\documentclass[twocolumn]{aastex631}

\usepackage{CJKutf8}

\newcommand{\fmh}{\mbox{H$_2$CO}}
\newcommand{\chcn}{\mbox{CH$_3$CN}}

\newcommand{\mjypbm}{\mbox{mJy\,beam$^{-1}$}}
\newcommand{\jypbm}{\mbox{Jy\,beam$^{-1}$}}
\newcommand{\kms}{\mbox{km\,s$^{-1}$}}

\newcommand{\cc}{\mbox{cm$^{-3}$}}

\newcommand{\msol}{\mbox{$M_\odot$}}

\usepackage{float}
\usepackage{csvsimple}
\usepackage{booktabs}
\usepackage{graphicx}
\usepackage{subfigure}
\usepackage{graphicx}
\usepackage{hyperref}

\begin{document}
\begin{CJK}{UTF8}{gbsn}
\title{ALMA Observations of Massive Clouds in the Central Molecular Zone: External-Pressure-Confined Dense Cores and Salpeter-like Core Mass Functions}

\author[0009-0005-4295-5010]{Zhenying Zhang (张朕荧)}
\affiliation{School of Physics and Astronomy, Yunnan University, Kunming 650091, People’s Republic of China}
\affiliation{Shanghai Astronomical Observatory, Chinese Academy of Sciences, 80 Nandan Road, Shanghai 200030, People’s Republic of China}

\correspondingauthor{Xing Lu}
\email{xinglu@shao.ac.cn}

\author[0000-0003-2619-9305]{Xing Lu (吕行)}
\affiliation{Shanghai Astronomical Observatory, Chinese Academy of Sciences, 80 Nandan Road, Shanghai 200030, People’s Republic of China}

\author[0000-0002-5286-2564]{Tie Liu (刘铁)}
\affiliation{Shanghai Astronomical Observatory, Chinese Academy of Sciences, 80 Nandan Road, Shanghai 200030, People’s Republic of China}

\author[0000-0003-2302-0613]{Sheng-Li Qin (秦胜利)}
\affiliation{School of Physics and Astronomy, Yunnan University, Kunming 650091, People’s Republic of China}

\author[0000-0001-6431-9633]{Adam Ginsburg}
\affiliation{Department of Astronomy, University of Florida, 211 Bryant Space Science Center P.O Box 112055 Gainesville, FL 32611-2055, USA}

\author[0000-0002-8691-4588]{Yu Cheng (程宇)}
\affiliation{National Astronomical Observatory of Japan, National Institutes of Natural Sciences, 2-21-1 Osawa, Mitaka, Tokyo 181-8588, Japan}

\author[0000-0003-2300-2626]{Hauyu Baobab Liu}
\affiliation{Department of Physics, National Sun Yat-Sen University, No. 70, Lien-Hai Road, Kaohsiung City 80424, Taiwan, R.O.C.}
\affiliation{Center of Astronomy and Gravitation, National Taiwan Normal University, Taipei 116, Taiwan}

\author[0000-0001-7330-8856]{Daniel~L.~Walker}
\affiliation{UK ALMA Regional Centre Node, Jodrell Bank Centre for Astrophysics, The University of Manchester, Manchester M13 9PL, UK}

\author[0000-0002-4154-4309]{Xindi Tang (汤新弟)}
\affiliation{Xinjiang Astronomical Observatory, 150 Science 1-Street, Urumqi, Xinjiang 830011, Peopleʼs Republic of China}
\affiliation{University of Chinese Academy of Sciences, Beijing, 100080, Peopleʼs Republic of China}
\affiliation{Key Laboratory of Radio Astronomy,Chinese Academy of Sciences, Urumqi, 830011, Peopleʼs Republic of China}
\affiliation{Xinjiang Key Laboratory of Radio Astrophysics,Urumqi, 830011, Peopleʼs Republic of China}

\author[0000-0003-1275-5251]{Shanghuo Li (李尚活)}
\affiliation{Max Planck Institute for Astronomy, K\"onigstuhl 17, D-69117 Heidelberg, Germany}

\author[0000-0003-2384-6589]{Qizhou Zhang}
\affiliation{Center for Astrophysics | Harvard \& Smithsonian, 60 Garden Street, Cambridge, MA 02138, USA}

\author[0000-0003-2133-4862]{Thushara Pillai}
\affiliation{Haystack Observatory, Massachusetts Institute of Technology, 99 Millstone Road, Westford, MA 01886, USA}

\author[0000-0002-5094-6393]{Jens Kauffmann}
\affiliation{Haystack Observatory, Massachusetts Institute of Technology, 99 Millstone Road, Westford, MA 01886, USA}

\author[0000-0002-6073-9320]{Cara Battersby}
\affiliation{Department of Physics, University of Connecticut, 196A Auditorium Road, Unit 3046 Storrs, CT 06269, USA}

\author[0000-0002-4707-8409]{Siyi Feng (冯思轶)} 
\affiliation{Department of Astronomy, Xiamen University, Zengcuo’an West Road, Xiamen, 361005, Peopleʼs Republic of China}

\author[0000-0002-8389-6695]{Suinan Zhang (张遂楠)}
\affiliation{Shanghai Astronomical Observatory, Chinese Academy of Sciences, 80 Nandan Road, Shanghai 200030, People’s Republic of China}

\author[0000-0002-2826-1902]{Qi-Lao Gu (顾琦烙)}
\affiliation{Shanghai Astronomical Observatory, Chinese Academy of Sciences, 80 Nandan Road, Shanghai 200030, People’s Republic of China}

\author[0000-0001-5950-1932]{Fengwei Xu (许峰玮)}
\affiliation{Kavli Institute for Astronomy and Astrophysics, Peking University, Beijing 100871, People's Republic of China}
\affiliation{I. Physikalisches Institut, Universität zu Köln, Zülpicher Str. 77, D-50937 Köln, Germany}
\affiliation{Department of Astronomy, School of Physics, Peking University, Beijing, 100871, People's Republic of China}

\author[0000-0001-9822-7817]{Wenyu Jiao (焦文裕)}
\affiliation{Shanghai Astronomical Observatory, Chinese Academy of Sciences, 80 Nandan Road, Shanghai 200030, People’s Republic of China}

\author{Xunchuan Liu (刘训川)}
\affiliation{Shanghai Astronomical Observatory, Chinese Academy of Sciences, 80 Nandan Road, Shanghai 200030, People’s Republic of China}

\author[0009-0009-8154-4205]{Li Chen (陈立)}
\affiliation{School of Physics and Astronomy, Yunnan University, Kunming 650091, People’s Republic of China}

\author[0000-0003-4506-3171]{Qiu-yi Luo (罗秋怡)}
\affiliation{Shanghai Astronomical Observatory, Chinese Academy of Sciences, 80 Nandan Road, Shanghai 200030, People’s Republic of China}
\affiliation{School of Astronomy and Space Sciences, University of Chinese Academy of Sciences, No.\ 19A Yuquan Road, Beijing 100049, People’s Republic of China}
\affiliation{Key Laboratory of Radio Astronomy and Technology, Chinese Academy of Sciences, A20 Datun Road, Chaoyang District, Beijing, 100101, People’s Republic of China}

\author[0000-0002-0786-7307]{Xiaofeng Mai (麦晓枫)}
\affiliation{Shanghai Astronomical Observatory, Chinese Academy of Sciences, 80 Nandan Road, Shanghai 200030, People’s Republic of China}
\affiliation{School of Astronomy and Space Sciences, University of Chinese Academy of Sciences, No.\ 19A Yuquan Road, Beijing 100049, People’s Republic of China}

\author[0009-0005-7028-0735]{Zi-yang Li (李紫杨)}
\affiliation{School of Physics and Astronomy, Yunnan University, Kunming 650091, People’s Republic of China}
\affiliation{Shanghai Astronomical Observatory, Chinese Academy of Sciences, 80 Nandan Road, Shanghai 200030, People’s Republic of China}

\author{Dongting Yang (杨东庭)}
\affiliation{School of Physics and Astronomy, Yunnan University, Kunming 650091, People’s Republic of China}

\author[0009-0004-3244-3508]{Xianjin Shen (沈先进)}
\affiliation{School of Physics and Astronomy, Yunnan University, Kunming 650091, People’s Republic of China}

\author[0000-0002-5789-7504]{Meizhu Liu (刘梅竹)}
\affiliation{Center for Astrophysics, Guangzhou University, Guangzhou 510006, People's Republic of China}

\author[0000-0003-3540-8746]{Zhiqiang Shen}
\affiliation{Shanghai Astronomical Observatory, Chinese Academy of Sciences, 80 Nandan Road, Shanghai 200030, People’s Republic of China}

\begin{abstract}
We present Atacama Large Millimeter/submillimeter Array (ALMA) Band~6 (1.3~mm) observations of dense cores in three massive molecular clouds within the Central Molecular Zone (CMZ) of the Milky Way, including the Dust Ridge cloud~e, Sgr~C, and the 20~\kms{} cloud, at a spatial resolution of 2000~au. Among the 834 cores identified from the 1.3~mm continuum, we constrain temperatures and linewidths of 253 cores using local thermodynamic equilibrium (LTE) methods to fit the \fmh{} and/or \chcn{} spectra. We determine their masses using the 1.3~mm dust continuum and derived temperatures, and then evaluate their virial parameters using the \fmh{} and/or \chcn{} linewidths and construct the core mass functions (CMFs). We find that the contribution of external pressure is crucial for the virial equilibrium of the dense cores in the three clouds, which contrasts with the environment in the Galactic disk where dense cores are already bound even without the contribution of external pressure. We also find that the CMFs show a Salpeter-like slope in the high-mass ($\gtrsim$3--6~\msol{}) end, a change from previous works with our new temperature estimates. Combined with the possible top-heavy initial mass functions (IMFs) in the CMZ, our result suggests that gas accretion and further fragmentation may play important roles in transforming the CMF to the IMF.
\end{abstract}


\section{Introduction} 
The central molecular zone (CMZ) of the Milky Way, located at the Galactic Center with a radius of approximately 300~pc, exhibits distinctive characteristics compared to the star formation environment in the solar neighborhood \citep{Henshaw2023}. There is molecular gas with a mass exceeding $10^7$~\msol{} \citep{Morris1996,Longmore2013,battersby2024}, an average gas number density of approximately $10^4$~\cc{} \citep{Paglione1998,Longmore2013}, a high gas temperature of 50--100~K \citep{Ao2013,Ginsburg2016,Krieger2017}, a strong magnetic field of about 1~mG \citep{Pillai2015,Lu2024}, and strong turbulence of Mach numbers $\mathcal{M}$ of around 30 \citep{Rathborne2014,Henshaw2016}. \cite{Barnes2017} found an incipient star formation rate (SFR) for the CMZ of about 0.09$\pm$0.02~\msol{}\,yr$^{-1}$, which is one order of magnitude lower than expected based on the dense gas-star formation relationship \citep{Longmore2013,Kauffmann2017,Lu2019ApJS}. A possible explanation for this could be that the clouds in the CMZ are in an early stage of evolution or that strong turbulence suppresses the collapsing and fragmentation of the clouds \citep{Kruijssen2014,Krumholz2017}.

The initial mass function (IMF) is the distribution of stellar masses of zero-age main sequence stars.
It is crucial for various fields in astrophysics, from star formation to galaxy evolution. It is believed to be universally applicable \citep{Kroupa2002,Bastian2010,matthews2014protostars}, although potential variations with respect to local environments have been found \citep[e.g.,][]{Hopkins2018,Li2023Nature}. 
The high-mass end of the IMF roughly follows a power law of the following form: $\frac{\text{d}N}{\text{d}\log M} \propto M^{-\alpha}$. \cite{Kroupa2002} fitted a power law index of $\alpha=1.35$. However, there is still controversy over the origin of the IMF and its dependence on environment. 

Stars form in dense cores \citep{Bergin2007}. Therefore, it has been suggested that the IMF is related to the core mass function (CMF), an analog of the IMF that describes the distribution of core masses in a star formation region 
\citep{Hennebelle2008,Offner2014,Ntormousi2019}. Numerical simulations have produced CMFs with a shape consistent with that of the IMF \citep{Klessen2000,Padoan2011}. Therefore, studying the CMF is crucial for exploring the origin of the IMF and investigating the early stages of star formation. Recent results from the ALMA-IMF large program \citep{motte2022alma,ginsburg2022alma,Pouteau2023} suggested that as molecular clouds evolve from quiescent to burst phases, the CMF may change from Salpeter-like to top-heavy, and then it might return to Salpeter-like as clouds approach the end of their star-forming phase. 

The CMZ is a specific target of interest for studying the relation between the CMF and IMF in extreme star forming environment. The IMF in young massive star clusters in the CMZ has been measured to be top-heavy, i.e., with an overpopulation of higher mass stars with respect to the `canonical' power-law form of $\frac{\text{d}N}{\text{d}\log M} \propto M^{-1.35}$ \citep{Husmann2012,Hosek2019}. Meanwhile, previous ALMA observations of dense cores in the CMZ clouds found signatures of top-heavy CMFs \citep[e.g.,][]{Lu2020ApJL}. However, there has been a dearth of constraints on the temperature of dense cores in the CMZ while estimating the core masses. Usually, a uniform core temperature (e.g., 20~K) is assumed, which could be biased and lead to large uncertainties in the CMFs \citep[see Appendix~D of][]{Lu2020ApJL}. Furthermore, the analysis of virial equilibrium of dense cores in the CMZ has been hindered because of a lack of spectral line observations that are able to measure the turbulent line width. Without the knowledge of virial states of cores, we cannot separate gravitationally bound cores that are deemed to collapse and form stars from unbound cores that do not form stars.

\begin{figure*}
\centering
\includegraphics[width=1\textwidth]{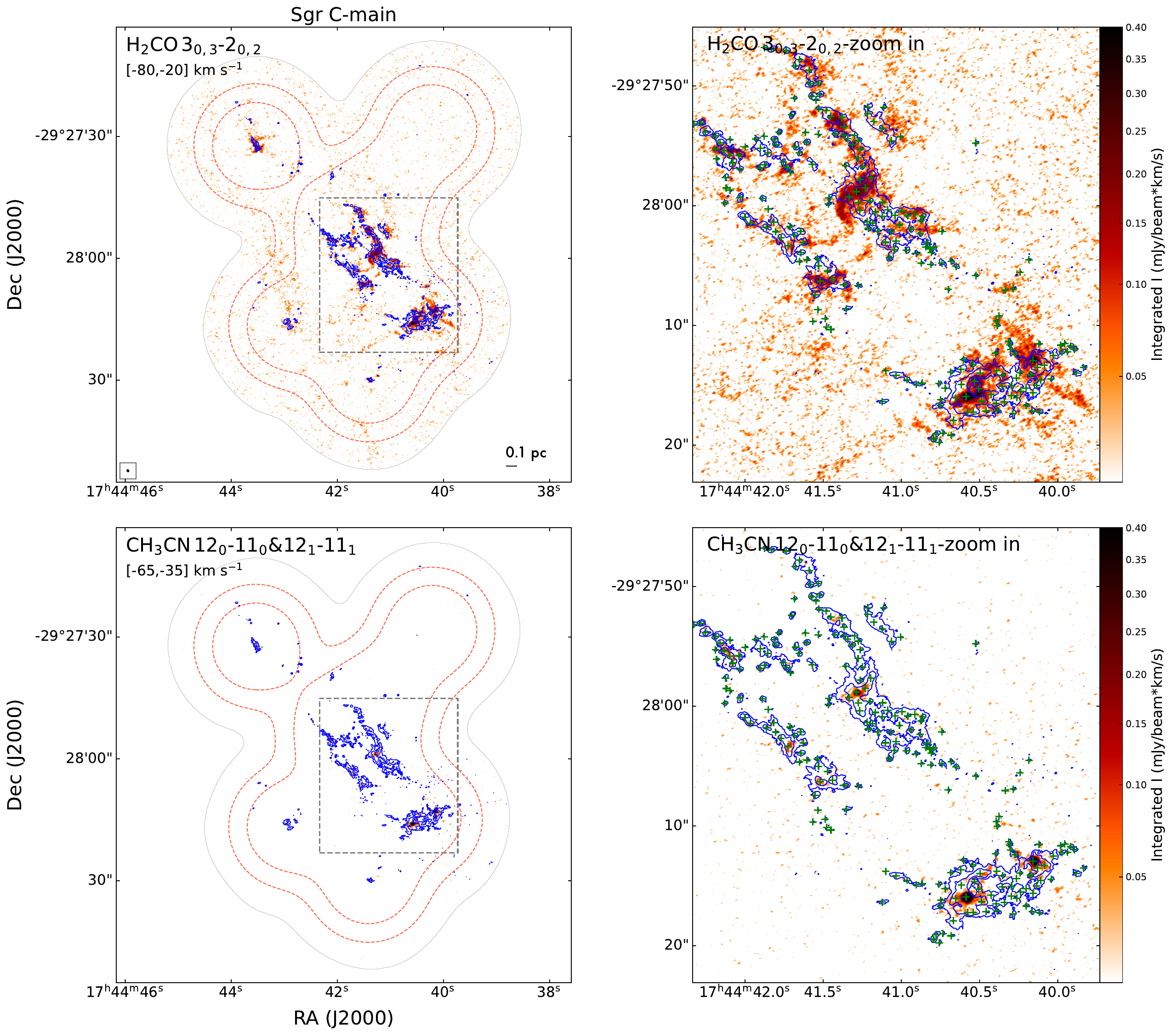}
\caption{The integrated intensity (0th moment) maps of the \fmh{} and \chcn{} line emission toward the three clouds. The inner and outer red dashed loops in the left panels show the ALMA primary-beam responses at 50\% and 30\%, respectively. The left panels show the integrated intensities of \fmh\,$3_{0,3}$--$2_{0,2}$ and \chcn\,$12_{0}$--$11_{0}$/ $12_{1}$--$11_{1}$, while the right panels are the zoomed-in views of the dashed boxes in the left panels. The blue contours show the 1.3 mm dust continuum at levels of 5 and 20 $\times \sigma$, where $\sigma=$  40~$\mu$\jypbm. The green `+' symbols illustrate the peak positions of the cores identified from the 1.3~mm continuum. (The complete figure set (7 images) is available in the online journal.)}
\label{fig:moment_0}
\end{figure*}

To more robustly characterize the masses and virial states of cores in the CMZ, we employ the \fmh{} and \chcn{} molecular lines to constrain the temperatures and velocity dispersions of cores in three massive molecular clouds in the CMZ. The rotational transitions of \fmh{} and the \chcn{} $K$-ladder are widely-used thermometers in observations of dense gas in clouds and cores \citep[e.g.,][]{Downes1980,Bieging1982,Zylka1992,Mangum2008,Ao2013,Tang2013}. In particular, we focus on the following \fmh{} transitions: $J_{K_1,K_2}=3_{0,3}$--$2_{0,2}$ ($E_{u}/k = 20.95$~K, $f_\mathrm{rest}=218.222192$~GHz), $2_{2,2}$--$2_{2,1}$ ($E_{u}/k = 68.09$~K, $f_\mathrm{rest}=218.47563$~GHz), and $2_{2,1}$--$2_{2,0}$ ($E_{u}/k = 68.11$~K, $f_\mathrm{rest}=218.76007$~GHz). Their critical densities at the temperature of 100~K is $\sim$$3\times10^{5}$~\cc{}. For \chcn{}, the transitions of interest include the $K$-ladder from $J_{K}=12_{0}$--$11_{0}$ ($E_{u}/k = 68.86$~K, $f_\mathrm{rest}=220.7472$~GHz) to $12_{7}$--$11_{7}$ ($E_{u}/k = 325.899$~K, $f_\mathrm{rest}=220.5944$~GHz), with a critical density at 100~K of $\sim$$5\times10^{6}$~\cc{}. The cloud sample includes the relatively quiescent Dust Ridge cloud~e (cloud~e hereafter), the 20~\kms{} cloud in an intermediate star formation state, and the actively star forming Sgr~C \citep{Lu2019ApJ,Lu2021apj}. These clouds are selected from a representative sample of massive clouds in the CMZ \citep{Kauffmann2017a,Kauffmann2017} and show clear fragmentation in our ALMA 1.3~mm continuum observations at a resolution of 2000~au \citep{Lu2020ApJL}. Additionally, we incorporate the impact of external pressure into the virial analysis, by using the SMA molecular line data from \citet{Lu2019ApJ}.

The structure of the paper is organized as follows: In \autoref{obs}, we describe the observations and data reduction. In \autoref{results}, we present the observational results. We then discuss the virial parameters, and the CMF within the molecular clouds in \autoref{discussion}. Finally, we summarize our findings in \autoref{conclusions}. Throughout the paper, we adopt a distance of 8.277~kpc to the Galactic Center \citep{GRAVITY2022} and assume that all the clouds under consideration are at this distance.

\section{Observations and data reduction}
\label{obs}
The ALMA observations have been presented in \citet{Lu2020ApJL,Lu2021apj}, where details of the observational setups and data reduction can be found. Here we reiterate the information that is related to our analysis in the current work.

The ALMA 12-m array observations in Band~6 toward the three clouds were conducted in the C40-5 and C40-3 configurations during April and July 2017 under the Project 2016.1.00243.S (PI: Q.\ Zhang). The correlators were configured to cover frequency ranges of 217--221~GHz and 231--235~GHz, with a uniform frequency resolution of 0.977~MHz (equivalent to a velocity resolution of 1.3~\kms). In this paper, we present the results derived from the 1.3~mm continuum and molecular line data of \fmh{} transitions $3_{0,3}$--$2_{0,2}$, $3_{2,2}$--$2_{2,1}$, and $3_{2,1}$--$2_{2,0}$, as well as the \chcn{} transitions $12_{0}$--$11_{0}$ to $12_{7}$--$11_{7}$.

The calibration and imaging of the data were performed using the Common Astronomy Software Application (CASA) version 5.4.0 \citep{CASATeam2022}. The calibrated data were then imaged using the \texttt{tclean} task with a robust Briggs parameter of 0.5. The achieved angular resolution of the images is approximately 0\farcs{24}$\times$0\farcs{18}. The sensitivity of the spectral line data is in the range of 1.6--2.0~\mjypbm{} per 1.3~\kms{} channel.

\section{RESULTS}
\label{results} 
In \citet{Lu2020ApJL}, the identification of dense cores in the three clouds has been carried out using the dendrogram algorithm implemented in the astrodendro package\footnote{\url{http://www.dendrograms.org}} upon the continuum images, where the `leaves' of the dendrograms were considered as dense cores. The three key parameters in astrodendro, the minimum intensity, the minimum significance, and the minimum area of leaves were set to 4$\sigma$, 1$\sigma$, and one synthesized beam size (i.e., 30.125 pixels), respectively, where $\sigma$ = 40~\mjypbm{}. In cloud~e, Sgr~C, and the 20~\kms{} cloud, 89, 274, and 471 cores were identified, respectively. The 1.3~mm continuum flux within the boundary of each leaf, defined by the 4$\sigma$ contour, without removing any background contribution from the larger structures within the dendrogram hierarchy, was adopted as the flux of a core. Here we directly adopted the core catalogs of \citet{Lu2020ApJL}, and extracted the \fmh{} and/or \chcn{} spectra toward the cores for fitting.

As discussed in \citet{Lu2020ApJL}, the maximum recoverable scale as determined by the short baselines in the ALMA array during our observations is 7\arcsec{} ($\sim$0.3~pc), whereas we focus on the dense cores at much smaller scales of 2000~AU. The missing flux issue of ALMA as an interferometer unlikely affects the measurement of fluxes of the cores significantly.

\begin{figure*}
\centering
\includegraphics[height=12cm]{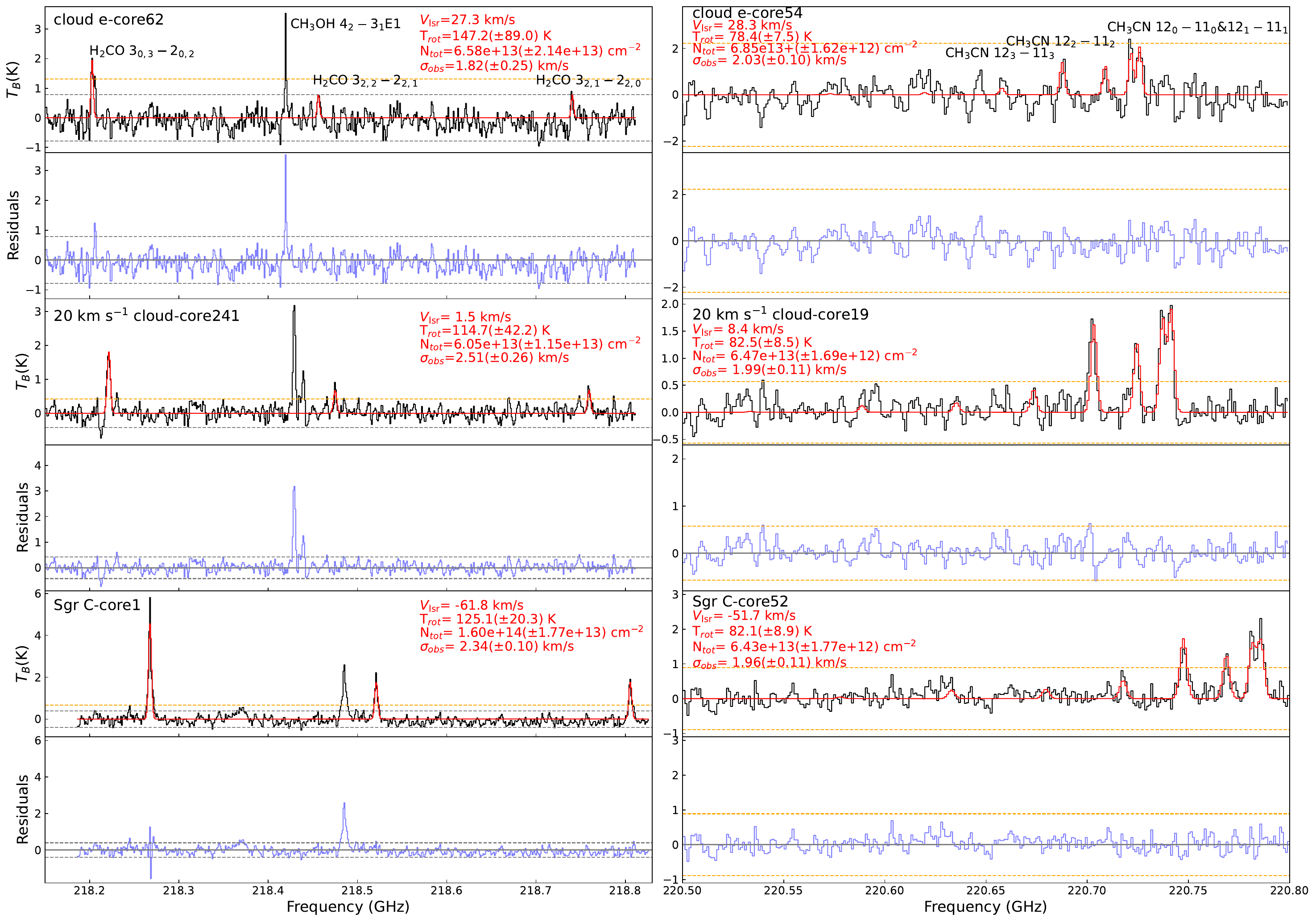}
\caption{Examples of \fmh{} (left) and \chcn{} (right) spectra detected in three cores, each from one of the three clouds. The complete gallery of all the fitting results can be found in \autoref{appendix A}. The black curves show the observed spectra, while the red curves are the results of the LTE fits. The gray dashed lines indicate the $\pm$3$\sigma$ levels, while the orange dashed line represents the 5$\sigma$ level. Above the spectra display the core indices, the best-fit centroid velocity, temperature, column density, and velocity dispersion. The blue curves below the spectra represent the fit residuals.}
\label{fig:example-H2CO/CH3CN}
\end{figure*}

\subsection{Overview of the spectral line emission}\label{subsec:results_lineemission}

\autoref{fig:moment_0} illustrates the integrated intensity maps of the \fmh\,$3_{0,3}$--$2_{0,2}$ and \chcn\,$12_{0}$--$11_{0}$/$12_{1}$--$11_{1}$ transitions in the three molecular clouds. Note that no masking was applied to the moment maps. The velocity ranges adopted for integrating the intensities are marked in the panels. We integrated the $12_{0}$--$11_{0}$ and $12_{1}$--$11_{1}$ transitions of \chcn{} altogether because the two lines are blended and therefore cannot be separated.

As shown in \autoref{fig:moment_0}, we found that the \chcn{} emission is mostly only detected towards the 1.3~mm emission.
However, the $\fmh\,3_{0,3}$--$2_{0,2}$ emission is not always associated with the 1.3~mm continuum, especially toward Sgr~C and the 20~\kms{} cloud where a significant number of filamentary structures in the \fmh{} emission are observed. A subset of these filaments have been identified as protostellar outflows in \citet{Lu2021apj}. The others are likely related to pc-scale shocks, which will be discussed in other works of the series (K.\ Yang et al.\ in prep.). For the \fmh{} emission spatially coincident with the identified cores, we assumed that it traces dense gas in the cores and therefore can be used to estimate the gas temperature and linewidth of the cores. In \autoref{appendix C}, we compared \fmh{} gas associated with the cores and that not associated with any cores, and found that the two samples of \fmh{} emissions present statistically different temperatures, with the \fmh{} associated with the cores showing lower temperatures.

\begin{figure*}[!t]
\centering
\includegraphics[width=0.49\textwidth]{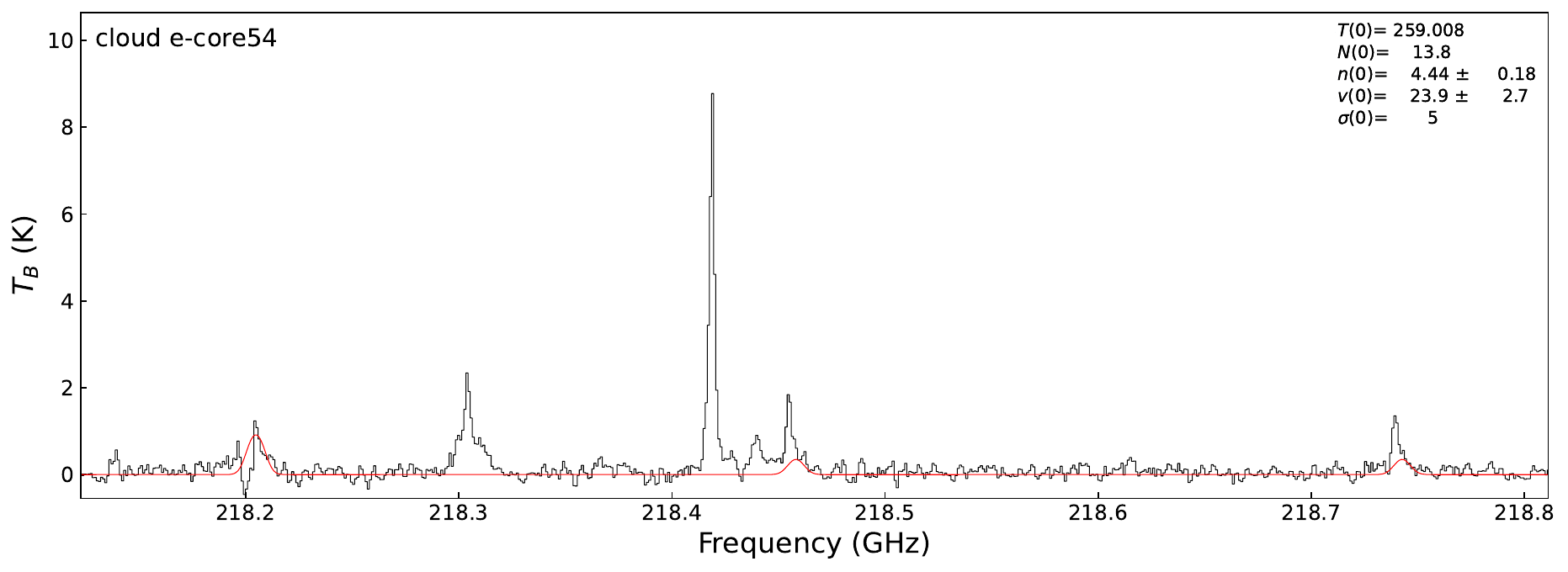}
\includegraphics[width=0.49\textwidth]{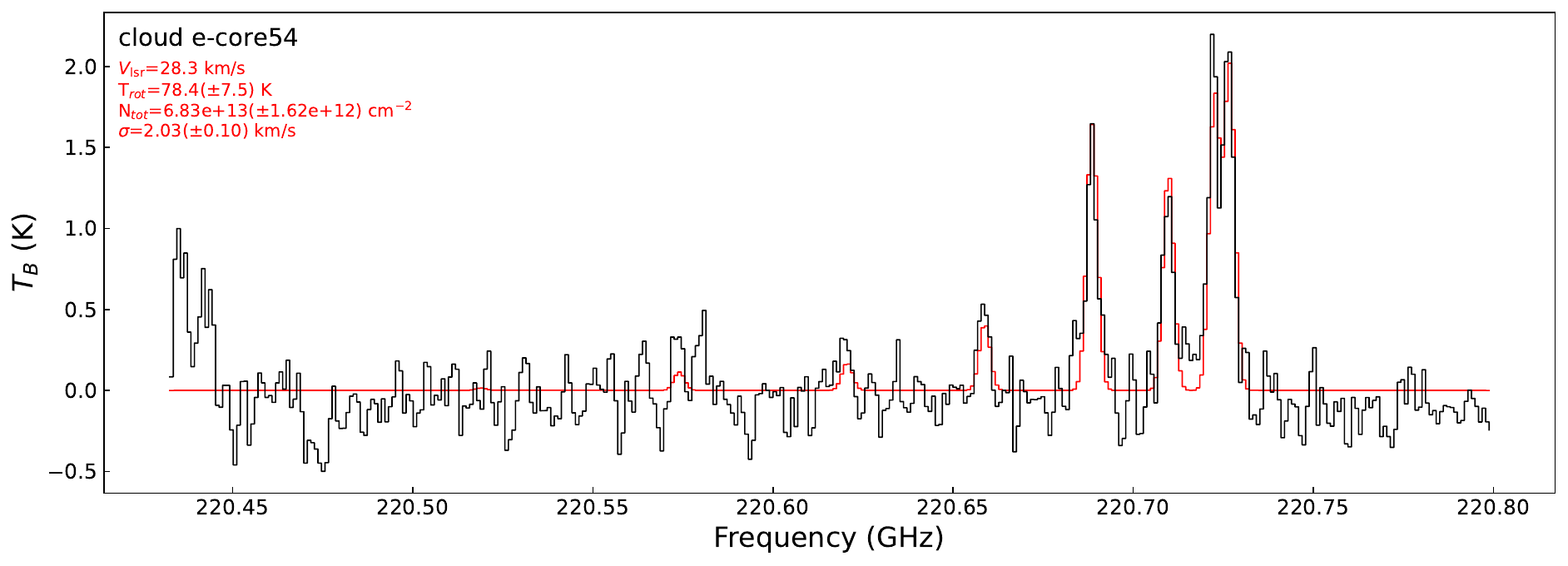}
\caption{The LTE fitting results for core 54 in cloud~e, using the FFTL code to fit \fmh{} (left) and the EMANON code to fit \chcn{} (right). The \fmh{} lines show clear self absorption, and therefore cannot be fit reasonably.}
\label{fig:self-absorption-example}
\end{figure*}

\begin{figure*}
\centering
\renewcommand\thefigure{\arabic{figure}} 
\includegraphics[height=13cm]{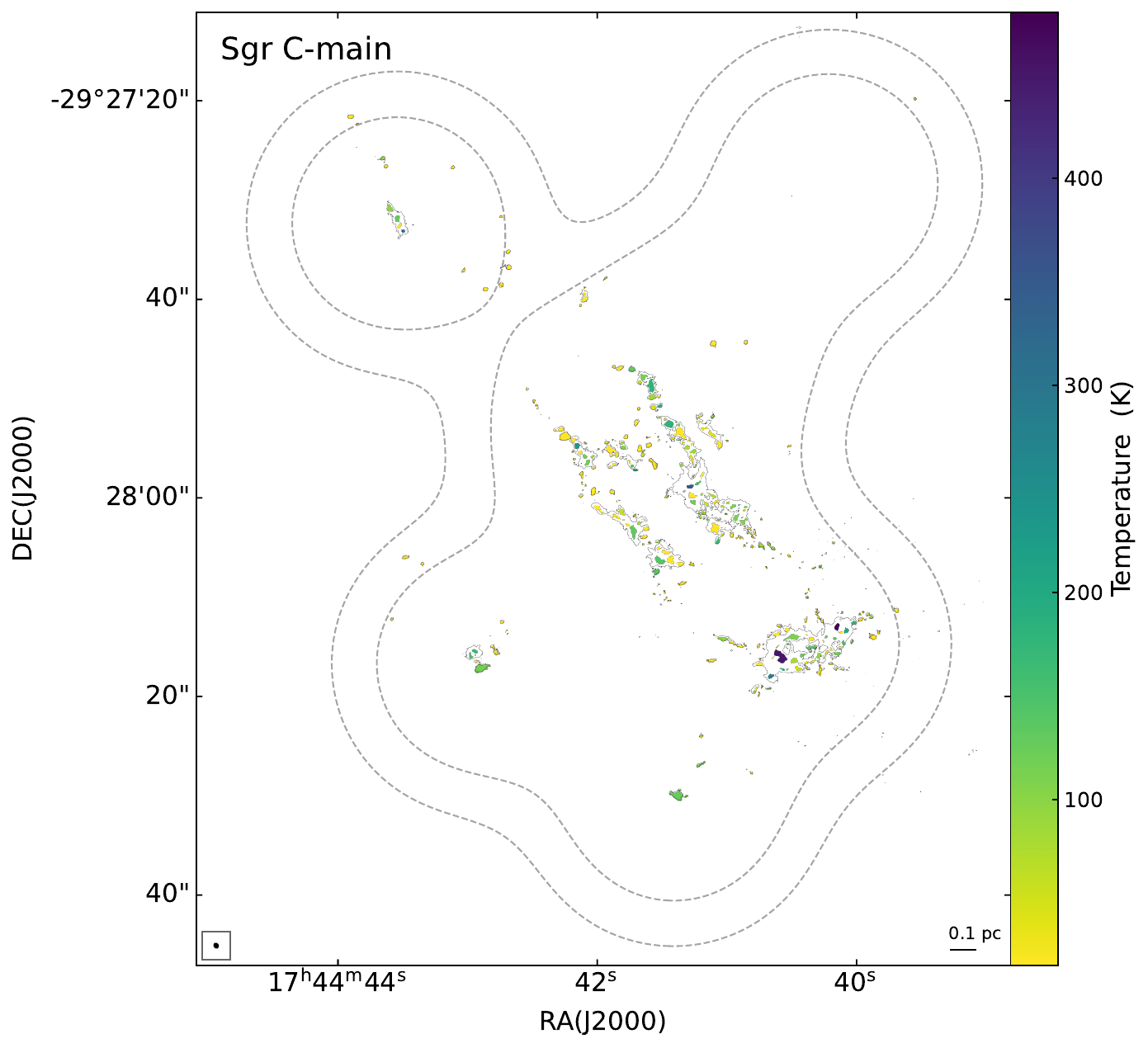}
\caption{An overview of the temperatures of the cores. The inner and outer dashed loops represent the ALMA primary-beam responses at 50\% and 30\%, respectively. The black contour represents the 1.3~mm continuum emission at the level of 5$\sigma$. The color patches show the positions and morphologies of the cores (see \autoref{tab:three cloud only 1o}), color-coded by their temperatures. Note that for cores (132/268) where neither \fmh{} nor \chcn{} is detected, we assume a temperature of 20~K. (The complete figure set (7 images) is available in the online journal.)}
\label{fig:allcores}
\end{figure*}

\subsection{Core temperatures from $\fmh$ and $\chcn$}\label{subsec:results_Tcore}

\autoref{fig:example-H2CO/CH3CN} left provides examples of the observed mean spectra of \fmh{} within dense cores from the three clouds, showing the three transitions \,$3_{0,3}$--$2_{0,2}$, $3_{2,2}$--$2_{2,1}$, and $3_{2,1}$--$2_{2,0}$. \autoref{fig:example-H2CO/CH3CN} right presents the observed mean spectra of \chcn{} within dense cores from the three clouds, encompassing lines from $12_{0}$--$11_{0}$ to $12_{7}$--$11_{7}$. Gaussian fits were employed on the mean \fmh{} and \chcn{} spectra to estimate temperatures and velocity dispersions of the dense cores, which were later used for estimating the masses and virial states of the cores.

Note that when both \fmh{} and \chcn{} lines are detected towards a core, the fitting result using \chcn{} was adopted, as in such cases the \fmh{} lines are always optically thick and show self absorption making the fit impractical (see \autoref{fig:self-absorption-example}).

\subsubsection{Fitting spectral lines}\label{subsubsec:results_fitting}
For the fitting of the \fmh{} lines, we employed the FFTL code\footnote{\url{https://github.com/xinglunju/FFTL}} that assumes the LTE conditions. The LTE is likely achieved in the cores given the high densities of $\gtrsim10^6$~\cc{} \citep{Goldsmith2001}. Nonetheless, to validate the results, we also performed fitting using the non-LTE method implemented with RADEX \citep{vandertak2007} within the PySpecKit package \citep{ginsburg2022pyspeckit}, with results presented in \autoref{appendix A}. We found consistent temperatures and velocity dispersions between the LTE and non-LTE methods as shown in \autoref{fig:error-T-sigma}.

The FFTL code performs forward modeling, in which model spectra are constructed based on the input temperature, column density, velocity dispersion, and the centroid velocity. Then the difference between the model spectra and the observed one is minimized using the \textit{lmfit} package\footnote{\url{https://lmfit.github.io/lmfit-py/}} to derive the best-fit model spectrum, whose input parameters (most notably, the temperature and the velocity dispersion) are taken as the final fitting result.

When selecting dense cores for fitting \fmh{}, we required the \fmh\,$3_{0,3}$--$2_{0,2}$ spectra to have a signal-to-noise (S/N) ratio higher than 5, where the RMS noise of the mean spectra was measured between the frequency range of 218.60--218.65~GHz that is mostly line-free. Note that there are several line-rich hot cores where emission of certain lines still exists in this frequency range. The RMS noise for these hot cores could be overestimated. Then there were two scenarios:

i) When one or both of the other two \fmh{} transitions, $3_{2,2}$--$2_{2,1}$ and $3_{2,1}$--$2_{2,0}$, have S/N ratios higher than 3, we fit the three transitions simultaneously using the FFTL code. 

ii) When both of the two transitions have S/N ratios lower than 3, we derived an upper limit for the temperature: we used the 3$\sigma$ value as the upper limit for the peak intensity of the two lines, and got the upper limit of the line ratio between these two transitions and the $3_{0,3}$--$2_{0,2}$ transition. This line ratio was then converted to a temperature upper limit following a best-fit power law relation between the line ratio and the temperature, which is derived from the best-fit temperatures in scenario i). There are 24 cases in the three clouds where we take this approach to estimate upper limits of the core temperature. Such upper limits are marked with `$<$' in front of the temperature values in \autoref{tab:three cloud only 1o}. Details of the line ratio-temperature relation can be found in \autoref{appendix B}.

For the cores where \chcn\,$12_{0}$--$11_{0}$/$12_{1}$--$11_{1}$ lines have a S/N ratio higher than 5, we also fit the \chcn{} $12_{0}$--$11_{0}$ to $12_{7}$--$11_{7}$ transitions to derive the temperature and velocity dispersion. The RMS noise was measured from the line-free channels in the frequency range of 220.55--220.58~GHz. We utilized the \textit{emanon} code\footnote{\url{https://github.com/xinglunju/emanon}, named after a fiction by Shinji Kajio.} with the LTE assumption, which adopts the same forward-fitting approach as FFTL.

As mentioned above, for all the cores with \chcn{} detections, the \fmh{} lines are also detected. We adopted the fitting results of \chcn{}, as the \fmh{} lines are optically thick and exhibit pronounced self-absorption, and therefore cannot be fit robustly (\autoref{fig:self-absorption-example}). We note that a systematic bias may exist between the best-fit parameters from the two lines, as they trace different gas components within the cores given their different excitation conditions. \chcn{} likely traces the hotter interiors of the cores, whereas \fmh{} traces more extended dense gas.

To validate the fitting results of \chcn{}, we conducted a cross-check using the XCLASS package that assumed the LTE conditions as well \citep{Moller2017}. As shown in \autoref{fig:error-T-sigma-CH3CN} of \autoref{appendix A}, we found consistent results between the \textit{emanon} code and the XCLASS package.

In the end, for cloud~e, Sgr~C, and the 20~\kms{} cloud, 15, 120, and 85 cores were successfully fitted with \fmh{}, and 4, 17, and 12 cores were fitted with \chcn{}. \autoref{fig:allcores} shows an overview of the core temperature. All the \fmh{} and \chcn{} spectra detected toward the cores and best-fit results are presented Figures~\ref{figA1} and \ref{figA2}. Histograms of temperatures and velocity dispersions are presented in \autoref{fig:T-linewidth}. The fitting results for the first 10 dense cores in each of the three clouds are showcased in \autoref{tab:three cloud only 1o}. A complete catalog of fitting results for all the dense cores within the three clouds can be found in \autoref{tab:cloud e table}, \autoref{tab:Sgr C cloud table}, and \autoref{tab:20km table}, respectively.

\subsection{Physical Properties of the Cores}
In \autoref{tab:three cloud only 1o}, physical properties of the dense cores within the three molecular clouds are provided. The catalog includes 19, 137, and 97 dense cores in cloud~e, Sgr~C, and the 20 \kms{} cloud, respectively, whose \fmh{} and/or \chcn{} spectra were successfully fitted.

\subsubsection{Core masses}
We assumed that gas and dust in the cores are coupled and therefore their temperatures are equal ($T_{\text{dust}}=T_{\text{gas}}$). This assumption are justified as the gas densities in the cores are expected to be sufficiently high \citep[$\geq$10$^6$~\cc{};][]{Goldsmith2001}: the densities were estimated to be $\sim$10$^7$~\cc{} assuming a $T_{\rm dust}$ of 20~K in \citet{Lu2020ApJL}; if higher temperatures of several hundreds K were adopted (see \autoref{subsec:results_Tcore}), the resulting densities would be an order of magnitude lower but still $\sim$10$^6$~\cc{}.

As discussed in \autoref{subsubsec:results_fitting}, when both \chcn{} and \fmh{} lines are detected towards a core, we adopted the best-fit LTE temperature of the former. When only \fmh{} is detected, the best-fit LTE temperature using the \fmh{} lines is adopted. If neither \chcn{} nor \fmh{} is detected, a dust temperature of 20~K is adopted, which is the assumption in \citet{Lu2020ApJL} and corresponds to the typical value of dust temperatures in cold cores \citep{Bergin2007}.

Assuming optically thin dust emission, the core mass was estimated following \citet{Hildebrand1983}:
\begin{equation}
 \label{eq:core_mass}
 M_{\text{core}} = \frac{D^2 S_\nu \eta}{\kappa_\nu B_\nu (T_{\text{dust}})},
\end{equation}
where $D$ is the distance to the core ($D$=8.277~kpc), $S_\nu$ is the dust emission flux, $\eta$=100 is the gas to dust mass ratio, $\kappa_\nu$=0.899~cm$^{2}$\, g$^{-1}$ \citep{Ossenkopf1994} is the dust absorption coefficient, $T_{\text{dust}}$ is the core temperature, and $B_{\nu}(T_{\text{dust}})$ is the Planck function at the temperature $T_{\text{dust}}$. Note that for the cores with only upper limits of temperatures (see \autoref{subsubsec:results_fitting}), the derived masses would be lower limits, which we have marked with `$>$' in front of their mass values in \autoref{tab:three cloud only 1o}. Also note that the gas-to-dust ratio in the CMZ could be as low as $\sim$50 \cite{Giannetti2017}. However, this does not affect our discussion of the CMF slope in \autoref{subsec:results_CMFs} where all core masses would scale down at the same rate, nor does it change the virial analysis in \autoref{subsubsec:results_virial}, where the virial parameters would increase by a factor of 2 while all the conclusions remain unaffected.

\subsubsection{Non-thermal motions in the cores}
We calculated the sound speed ($c_s$) and the Mach number ($\mathcal{M}$) following
\begin{equation}\label{eq:as}
c_s = \sqrt{\frac{k_{\text{B}}T_{\text{core}}}{\mu m_{\text{H}}}},
\end{equation}
and
\begin{equation}\label{eq:mach-number}
\mathcal{M} = \frac{\sigma_{\text{NT}}}{c_s}  = \frac{\sqrt{\sigma_{\text{tot}}^2 - c_s^2}}{c_s},
\end{equation}
where $k_{\text{B}}$ is the Boltzmann constant, $T_{\text{core}}$ the temperature of the core derived in \autoref{subsec:results_Tcore}, $\mu$ is the mean molecular weight and is taken to be 2.37 \citep{Kauffmann2008}, $m_{\text{H}}$ is the mass of the hydrogen atom, $\sigma_{\text{NT}}$ is the non-thermal velocity dispersion, and $\sigma_{\text{tot}}$ is the total (thermal and non-thermal) velocity dispersion. 

The $\fmh{}$ and $\chcn{}$ lines were fitted to estimate the velocity dispersion $\sigma_{\text{obs}}$, as described in \autoref{subsubsec:results_fitting}. Then, the channel width of $V_{\text{ch}}=1.35$~\kms{} was subtracted quadratically to derive the deconvolved velocity dispersion:
\begin{equation}
 \label{eq:three-dimensional}
 \sigma_{\text{deconv}}^2=\sigma^2_{\text{obs}} - \left(\frac{V_{\text{ch}}}{2\sqrt{2\ln{2}}}\right)^2.
\end{equation} 
The total velocity dispersion, which includes the thermal and non-thermal components, was then derived following
\begin{equation}
 \label{eq:tot-dimensional}
 \sigma^2_{\text{tot}} = \sigma^2_{\text{NT}} + c_s^2 = \left(\sigma_{\text{deconv}}^2 - \frac{k_{\text{B}} T_{\text{core}}}{\mu_{\text{mol}} m_{\text{H}}}\right) + \frac{k_{\text{B}} T_{\text{core}}}{\mu m_{\text{H}}},
\end{equation}
where the molecular weight $\mu_{\text{mol}}$ is 28 for \fmh{} and 44 for \chcn{}.

\begin{figure*}
\centering
\includegraphics[height=6cm]{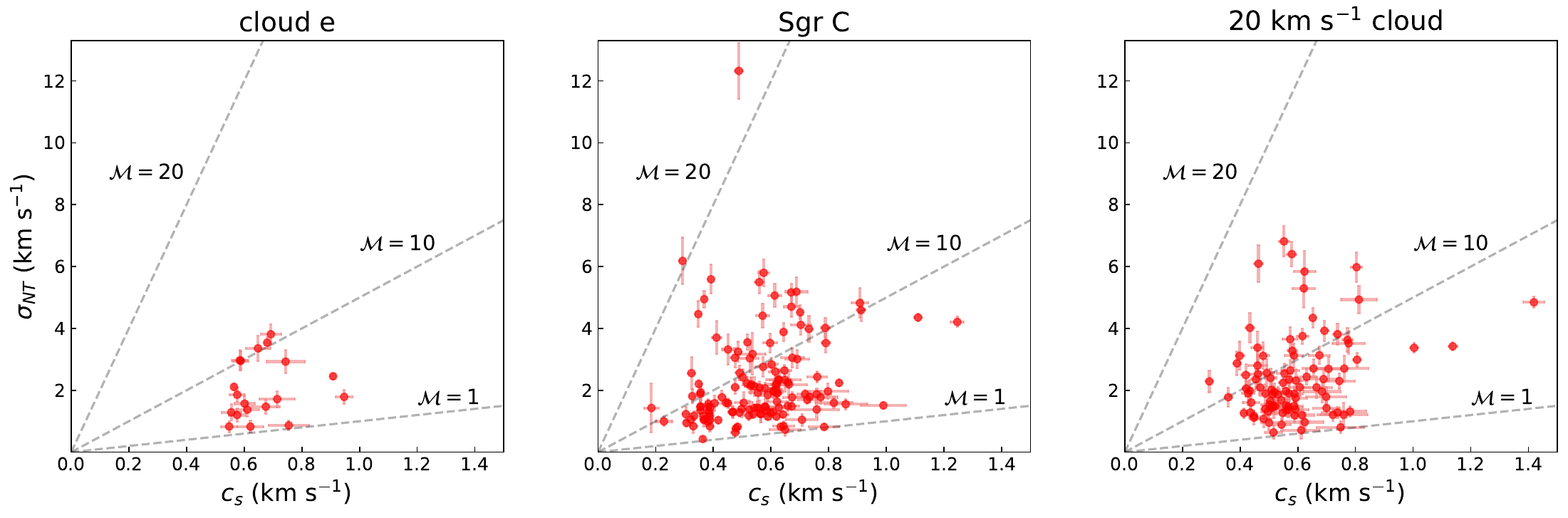}
 \caption{Comparisons between the non-thermal velocity dispersion ($\sigma_{NT}$) and the sound speed ($c_s$). The dashed gray lines represent Mach numbers $\mathcal{M} =$ 1, 10, and 20.}
\label{fig:m}
\end{figure*}

\autoref{fig:m} shows the relation between the non-thermal velocity dispersion ($\sigma_{\text{NT}}$) and sound speed ($c_s$) of the cores in the three clouds. The Mach numbers of the cores are listed in \autoref{tab:three cloud only 1o}, with their errors estimated from the error propagation. The mean Mach numbers in cloud~e, Sgr~C, and the 20~\kms{} cloud are 3.2, 4.4 and 4.3, respectively, suggesting supersonic motions in the cores. The result confirms that the cores in the three clouds are subject to supersonic non-thermal motions (e.g., turbulence, infall, or rotation), which is the typical condition for gas in the CMZ \citep[e.g.,][]{Rathborne2014}.

\subsubsection{Virial parameters}\label{subsubsec:results_virial}
As shown in \autoref{subsec:results_Tcore}, 253 dense cores have sufficiently high S/N ratios in \fmh{} or \chcn{} to estimate temperatures and velocity dispersions. We then performed virial analyses to assess dynamical states of these cores. We considered the contributions of kinetic energy ($\Omega_K$), gravitational potential energy ($\Omega_G$), and external pressure ($\Omega_P$) from surrounding gas in the virial equilibrium analysis.

Assuming a spherically symmetric Gaussian density distribution for the gravitational potential energy, the gravitational term ($\Omega_G$) can be derived following \citep{Bertoldi1992}:
\begin{equation}
 \label{eq:W_$G$}
 \Omega_{G} = -\frac{3}{5} a \frac{GM^2_{\text{core}}}{R},
\end{equation}
where $R$ is the effective core radius. We took the core area $A_{\text{core}}$ from astrodendro and derived $R$ following $R=\sqrt{A_{\text{core}}/\pi}$. The parameter $a$ is a constant whose value depends on the power-law index of the radial density distribution $\rho \propto r^{-k_p}$, $a=(1-k_p/3)/(1-2 k_p/5)$. Here we took $k_p=2$ and $a=5/3$, which is the case for self-gravitating cores \citep{Bertoldi1992}.

The kinetic energy of a core of mass $M_{\text{core}}$ and one-dimensional (1D) velocity dispersion $\sigma_{\text{tot}}$ is given by
\begin{equation}
\label{eq:$W_k$}
\Omega_{k} = \frac{3}{2} M_{\text{core}} \sigma^2_{\text{tot}}.
\end{equation}

Lastly, following e.g.\ \citet{Kirk2017} and \citet{Scibelli2023}, the external pressure term $\Omega_{P}$ is given by
\begin{equation}
 \label{eq:WP}
 \Omega_{P} = -4 \pi P_{\text{out}} R^3.
\end{equation}
$P_{\text{out}}$ is external pressure given by
\begin{equation}
 \label{eq:P}
 P_{\text{out}} = \rho_{\text{out}} \sigma^2_{\text{tot,out}} = \mu m_{\text{H}} n_{\text{out}} \sigma^2_{\text{tot,out}},
\end{equation}
where $\sigma_{\text{tot,out}}$ is thermal plus non-thermal velocity dispersion of the gas surrounding the core. We assumed that the cores are enveloped by gas at larger scales, as captured by the SMA observations at a resolution of 4$''$ in \citet{Lu2019ApJ}, and adopted the gas density $n_{\text{out}}$ estimated using the SMA 1.3~mm continuum (assuming a dust temperature of 20~K) and the velocity dispersion $\sigma_{\text{tot,out}}$ from the SMA N$_2$H$^+$ or CH$_3$OH lines \citep{Lu2019ApJ}. The adopted values for the cores are listed in \autoref{tab:three cloud only 1o}. Note that the SMA observations also suffered from the missing flux issue and filtered out spatially extended emission. However, such emission should represent more diffuse gas enclosing the gas component captured by the SMA, and thus not directly exerting pressure on the even small cores in the ALMA observations.

\begin{figure*}
\centering
\includegraphics[height=9cm]{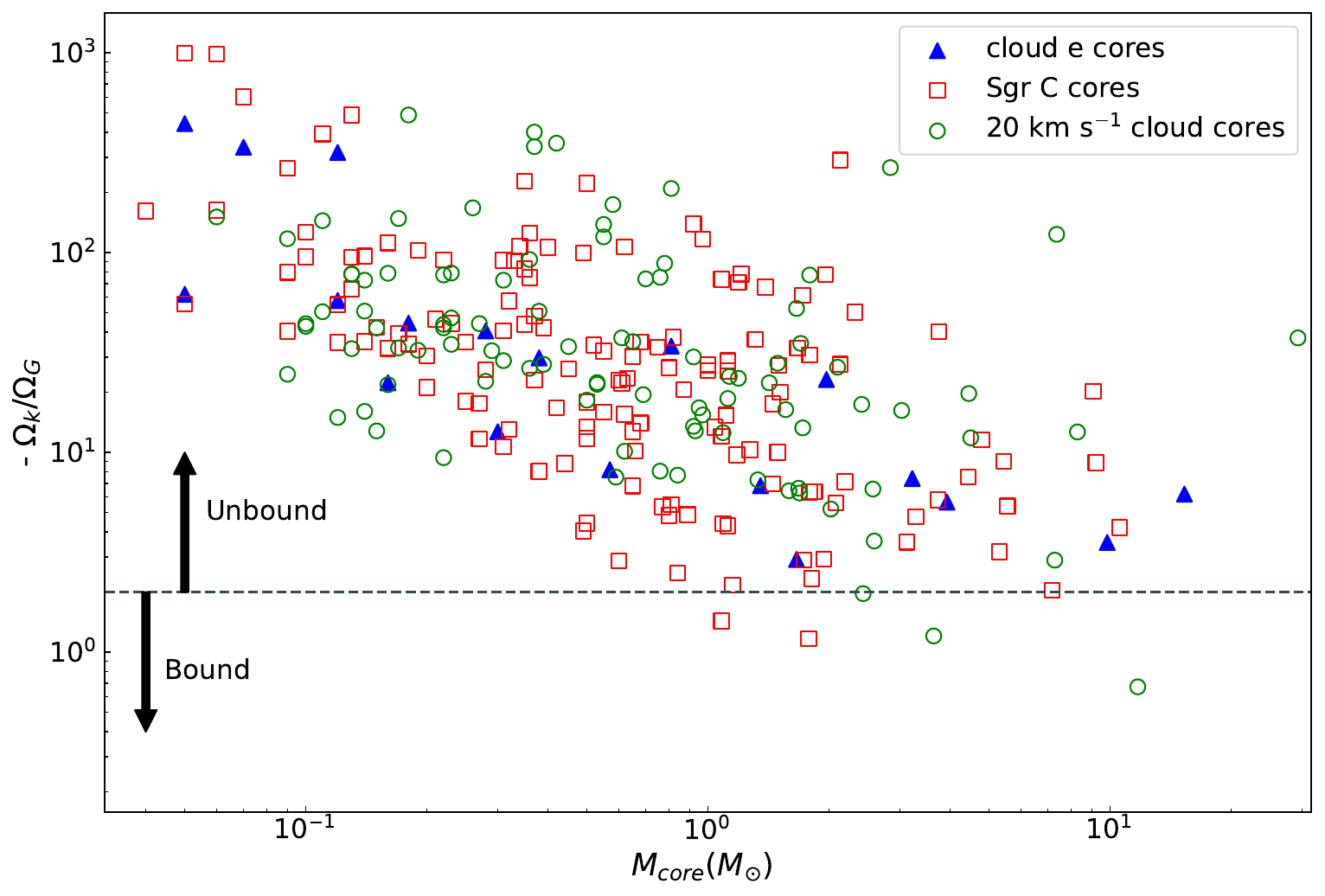}
\caption{Distribution of masses and virial parameters of the cores, considering only gravitational and kinetic energies. The data points are color-coded by the clouds they belong to. Cores located above the dashed line $(\alpha_{\rm vir} = -\Omega_k/\Omega_G =2)$ are considered to be unbound.}
\label{fig:virial-relation1}
\end{figure*}

We considered the equilibrium between gravitational and kinetic energies
\begin{equation}
 \label{eq:virial-no_P}
 2\Omega_k + \Omega_G = 0,
\end{equation}
and derived the virial parameter $\alpha_\mathrm{vir}$ as
\begin{equation}
 \label{eq:virial}
 \alpha_\mathrm{vir} = -\Omega_k/\Omega_G.
\end{equation}
If $\alpha_\mathrm{vir}$ is lower than the critical value of 2, a core would be gravitationally bound, otherwise it would be unbound \citep{Kauffmann2013}. \autoref{fig:virial-relation1} depicts the distribution of the virial parameters of the dense cores within the three clouds, which the dashed horizontal line marks the critical value of 2. It can be seen that the majority of the dense cores (247/253) are in an unbound state.

\begin{figure*}
\centering
\includegraphics[height=9cm]{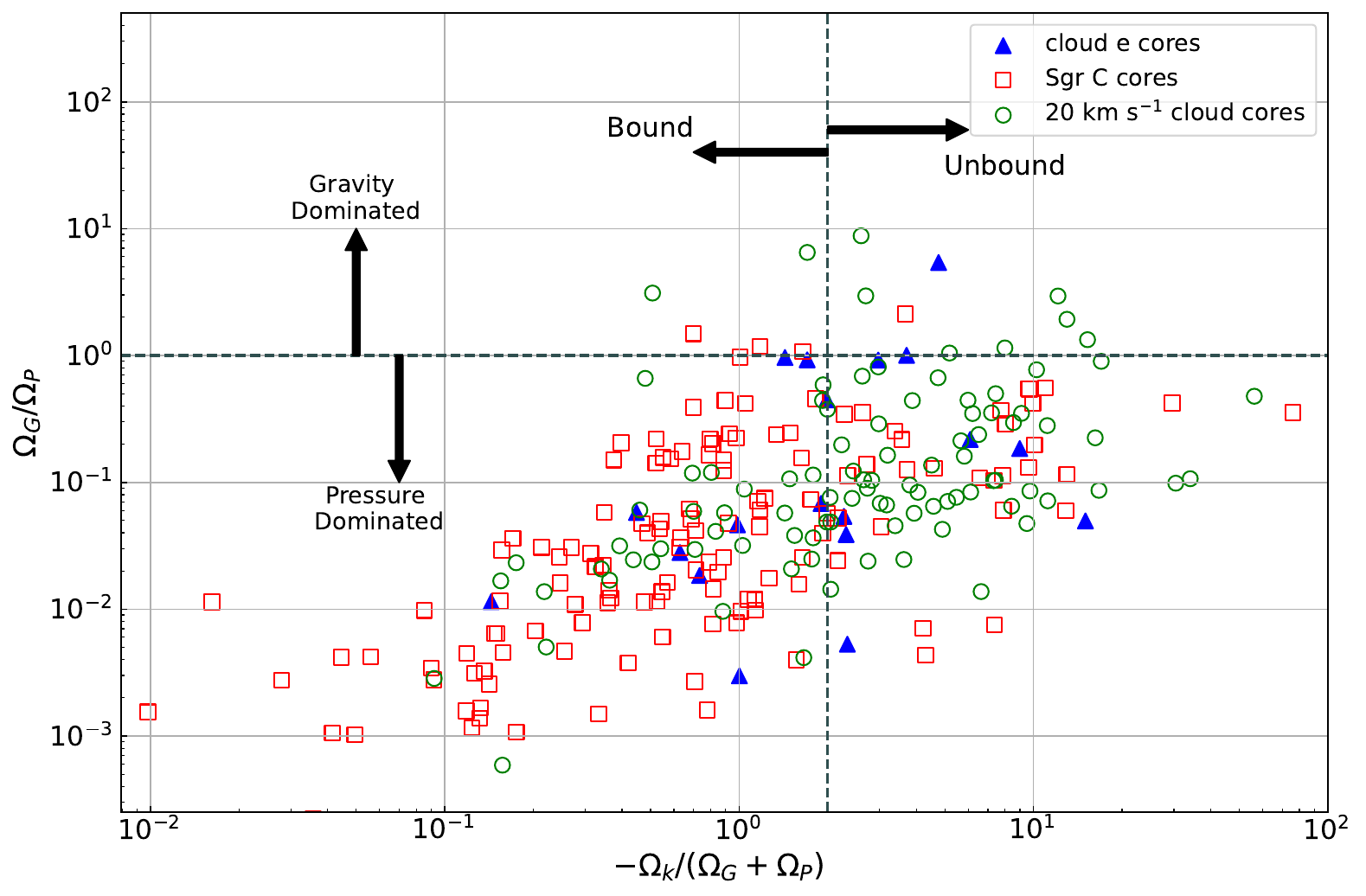}
\caption{Virial parameters ($\alpha_\mathrm{vir,p}$) versus confinement ratios ($\Omega_G/\Omega_P$). Cores located above the horizontal dashed line $( \Omega_G/\Omega_P =1)$ are dominantly confined by gravity, while those below are dominantly confined by external pressure. Cores located to the left of the vertical dashed line $\alpha_\mathrm{vir,p} = -\Omega_k/(\Omega_G+\Omega_P)=2$ are considered to be bound, while those to the right are unbound.}
\label{fig:virial-relation2}
\end{figure*}

\begin{figure*}
\centering
\includegraphics[width=1\textwidth]{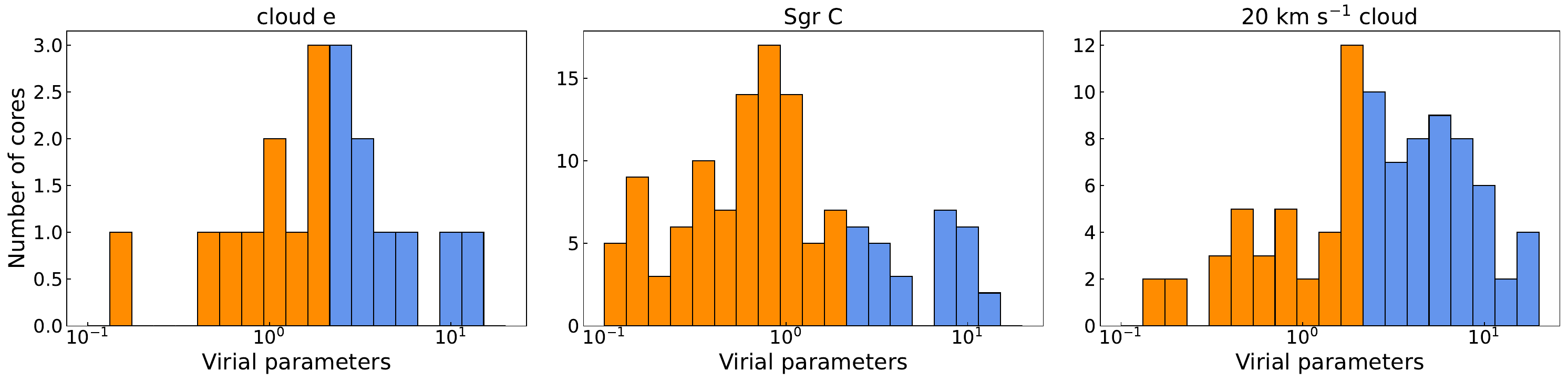}
\caption{Histograms of virial parameters of the cores. The orange bars represent the portion of cores with virial parameter \( \alpha_\mathrm{vir,p}=-\Omega_k/(\Omega_G+\Omega_p) < 2 \), which are 10/19, 108/137, 36/97 in the cloud~e, Sgr~C, and the 20 \kms{} cloud, respectively.}
\label{fig:virial_histogram}
\end{figure*}

If we additionally considered the contribution of external pressure
\begin{equation}
 \label{eq:virial-P}
 2\Omega_k + \Omega_G + \Omega_P = 0,
\end{equation}
and then derived the virial parameter as
\begin{equation}
 \label{eq:alpha}
 \alpha_\mathrm{vir,p} = -\frac{\Omega_k}{\Omega_G + \Omega_P}.
\end{equation}
We listed the virial parameters of the cores that take external pressure into account in \autoref{tab:three cloud only 1o}, and plotted the results in \autoref{fig:virial-relation2} (cf.\ \citealt{Pattle2015}; \citealt{Kirk2017}). Histograms of the virial parameters in the three individual clouds are shown in \autoref{fig:virial_histogram}.

As shown in \autoref{fig:virial-relation2}, taking the external pressure into account in the virial equilibrium leads to a greater number of cores being bound (154/253), including 10/19, 108/137, and 36/97 cores in cloud~e, Sgr~C, and the 20~\kms{} cloud, respectively, characterized by $\alpha_\mathrm{vir,p} < 2$. Moreover, the majority of the cores (239/253) have $\Omega_G/\Omega_P < 1$, indicating that the cores are predominantly confined by external pressure instead of self gravity, including 18/19, 134/137, 88/97 cores in cloud~e, Sgr~C, and the 20~\kms{} cloud, respectively. 

Our finding contrasts with results toward clouds in the Galactic disk, where most of the cores are already bound even without considering the contribution of external pressure in the virial equilibrium \citep[e.g.,][]{Kauffmann2013,lis2023,cheng2024}, and reveals that non-thermal motions in the form of external pressure and internal support dominate gas dynamics in the spatial scale of individual cores ($\sim$2000~au). This result echoes with previous studies that highlight high pressure star forming environments in the CMZ on spatial scales of $\sim$0.1~pc \citep{Liu2013,Walker2018,Myers2022,Callanan2023} and $\sim$10$^3$~AU \citep{Williams2022}.

\subsection{Core Mass Functions}\label{subsec:results_CMFs}

To fit a power law to the high-mass end of the CMF, we employed the maximum likelihood estimation (MLE) method \citep{Clauset2009} implemented in the \textit{plfit} package\footnote{\url{https://github.com/keflavich/plfit}}, following the form
\begin{equation}
 \label{eq:ha1}
 \frac{{\rm d}N}{{\rm d}\log M} \propto M^{-\alpha}.
\end{equation} 
The power-law index ($\alpha$) and the lower limit of the power law relation ($M_\mathrm{min}$) were simultaneously determined during the fitting \citep[see e.g.,][]{Lu2020ApJL}.

\begin{figure*}
\centering
\includegraphics[height=12cm]{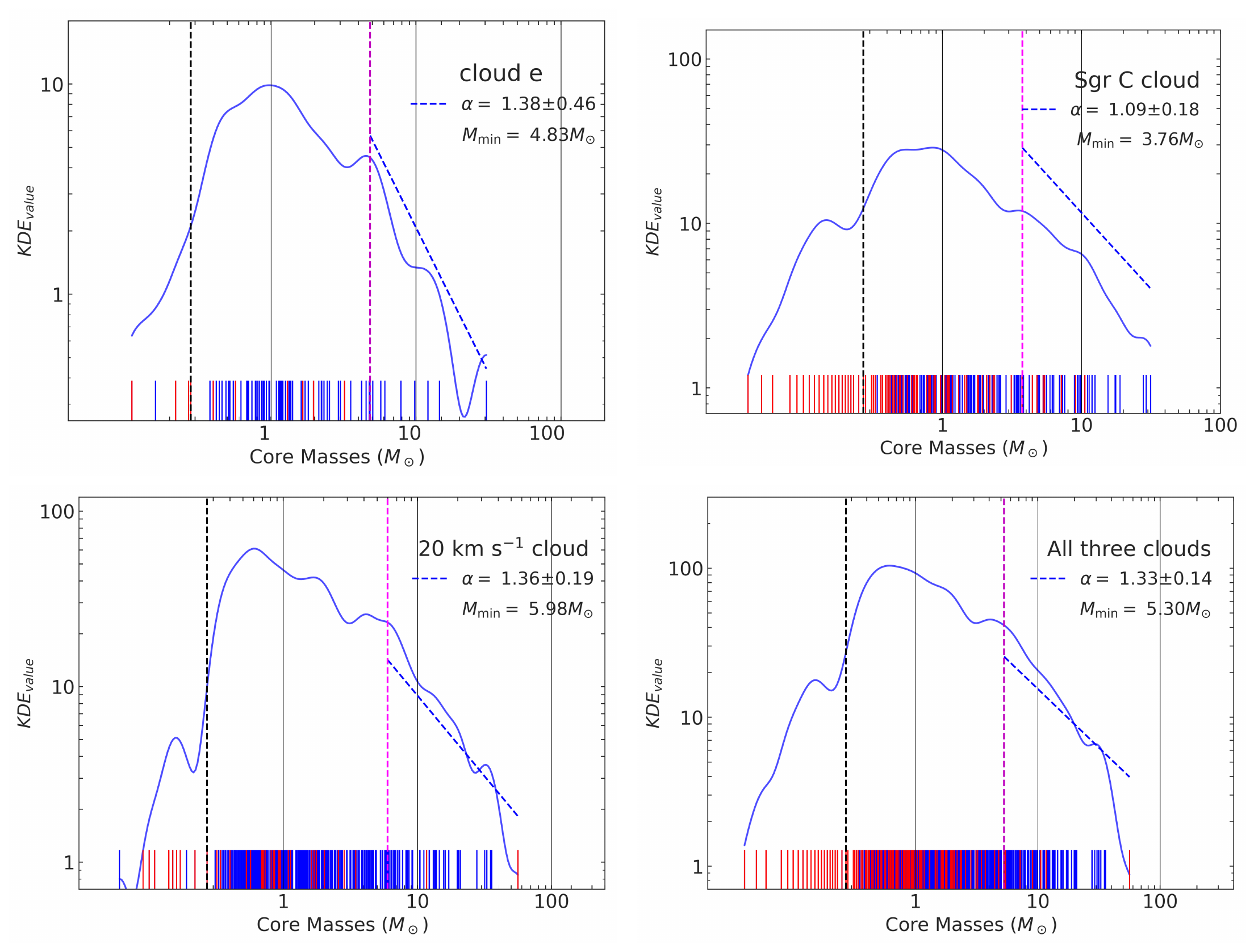}
 \caption{CMFs for the three individual clouds and the three clouds combined. The blue curves represent the kernel density estimations (KDEs) of the core masses. The red sticks attached to the bottom horizontal axes denote the core masses obtained with best-fit temperatures from \fmh{} or \chcn{}, while the blue ones represent the core masses assuming a dust temperature of 20~K. The black vertical dashed lines represent the 5$\sigma$ mass sensitivity (0.27~\msol) at a dust temperature of 20~K. The magenta vertical dashed lines represent the minimum core masses of the power-law fittings as determined by \textit{plfit}, and the slanted blue dashed lines represent the best-fit power-law relations with an arbitrary normalization. Note that we choose not to plot histograms of core masses, because we do not bin the masses to fit the power-law, but carry out the power-law fitting using all the actual data above the minimum core masses following the MLE method.}
\label{fig:cmf}
\end{figure*}

We fit the CMFs for each cloud as well as for the three clouds combined, as shown in \autoref{fig:cmf}. Note that we have excluded those sources with $\alpha_\mathrm{vir,p}>2$ and therefore being unbound. In the end, 80, 245, and 410 cores were considered for cloud~e, Sgr~C, and the 20 \kms{} cloud, respectively. The best-fit power-law indices $\alpha$ for each individual cloud are between 1.09 and 1.38. For the combined CMF of the three clouds, the best-fit $\alpha$ is 1.33$\pm$0.14. These power-law indices are higher than those reported in \citet{Lu2020ApJL}, which were derived from CMFs of the same sample of dense cores, yet with an assumed universal dust temperature of 20~K. In \autoref{appendix D}, we conducted several tests, e.g., changing the assumed temperatures for the cores without \fmh{}/\chcn{} line detections from 20~K to 30~K or 50~K, using the temperatures derived from the non-LTE method, and only using cores with temperatures estiamted from the \fmh{}/\chcn{} lines, and found that the CMFs consistently show higher power-law indices than those in \citet{Lu2020ApJL}.

\autoref{fig:KS-test} displays the cumulative distribution functions (CDFs) of core masses derived from a constant dust temperatures of 20~K \citep{Lu2020ApJL} and from the updated core temperatures in the current work. The $p$-values of two-sample Kolmogorov-Smirnov (K-S) tests between the two CDFs are 0.63, $2.9\times10^{-7}$, and 0.01, for the three clouds respectively, and $2.9\times10^{-9}$ for the three clouds combined. Usually when $p < 0.05$, one considers that the two samples in the K-S test are not drawn from the same distribution. As such, the core masses estimated with the updated temperatures do represent a statistically different distribution than those reported in \citet{Lu2020ApJL}, except for cloud~e where the number of cores is lower and therefore the statistics may not be robust.

\begin{figure*}
\centering
\includegraphics[width=1\textwidth]{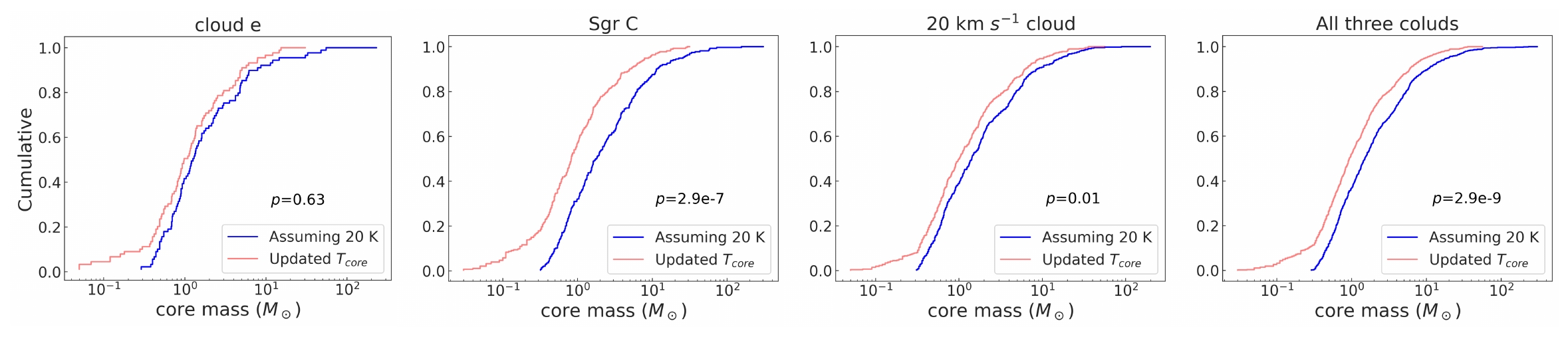}
\caption{Cumulative distributions of the core masses in the current work and from \citet{Lu2020ApJL}. The blue curves represent the cumulative mass distributions assuming a dust temperature of 20~K for all cores. The red curves represent the mass distributions with partially updated temperatures from fitting \fmh{} and \chcn{} lines. The $p$-values of the two-sample K-S tests between the cumulative distributions are displayed in the bottom-right corners of the panels.}
\label{fig:KS-test}
\end{figure*}

We noted that the estimated power-law indices of the high-mass ends of the CMFs are consistent with that of the canonical IMF with $\alpha$=1.35 \citep{Kroupa2002}. That is, the CMFs of the three clouds exhibit a Salpeter-like shape in their high-mass ends. The implications of these findings will be discussed in \autoref{discussion-2}.

\section{Discussion}
\label{discussion}

\subsection{Dynamical states of the cores in different clouds}
\label{discussion-1}
\autoref{fig:virial_histogram} shows histograms of the virial parameters in the three clouds. The orange bars represent cores with virial parameters $\alpha_\mathrm{vir,p}<2$, which are expected to be bound. Toward the three clouds, there are 10, 108, and 37 cores that fall within this bound state. These cores will likely collapse and form stars.

Meanwhile, the blue bars indicate that a large population of cores are unbound. As seen in Figures~\ref{fig:virial-relation2} \& \ref{fig:virial_histogram}, even after considering the contribution of external pressure, there are still 9 out of the 19 cores in cloud~e, 29 out of the 137 cores in Sgr~C, and 61 out of the 97 cores in the 20~\kms{} with $\alpha_\mathrm{vir,p}>2$. The situation is more pronounced in the 20~\kms{} cloud compared to Sgr~C, while in cloud~e there are not statistically enough cores with molecular line detection to reach a conclusion.

The different fractions of bound or unbound cores in Sgr~C and the 20~\kms{} cloud may reflect their different evolutionary phases regarding star formation, or different cloud-scale dynamics when they orbit around the Galactic Center. To discriminate between these scenarios or explore other alternatives, a large sample of clouds in the CMZ at various evolutionary phases and orbital positions is needed to determine the viral states of dense cores. Our observations tentatively lend stronger support to the latter case:

\begin{itemize}
 \item The estimate of the virial parameter is clearly affected by the measurement of velocity dispersions. When a core is associated with outflows, the velocity dispersion is likely to be overestimated because outflows broaden the \fmh{} linewidth \citep[\fmh{} has been found to trace outflows in these clouds;][]{Lu2021apj}. Additionally, the presence of hot molecular cores, which are an advanced evolutionary phase of high-mass star formation, will also broaden the observed linewidth due to high temperatures (up to a few $10^2$~K) and strong feedback. In such cases, the higher ratios of unbound to bound cores in the 20 \kms{} cloud might suggest that star formation activities in the 20 \kms{} cloud are more active than in Sgr~C. However, this appears inconsistent with our observations. The number of outflows in the 20 \kms{} cloud and Sgr~C are 20 vs.\ 18, respectively, while the numbers of hot molecular cores (represented by cores with \chcn{} emission identified in the current work) are 12 vs.\ 17, respectively. These numbers are not significantly different. We conducted a test and found that after excluding cores associated with outflows and hot cores, the distribution of their virial parameters remained largely unchanged. 

 \item The different positions and evolutionary stages of the two clouds in the CMZ may lead to different cloud-scale dynamics. The 20~\kms{} cloud is closer to the center of gravitational potential in most of the 3D models of the CMZ \citep[see Section 4.3.2 of][]{Henshaw2023} and therefore is expected to experience stronger tidal effects than Sgr~C. In certain models \citep[e.g.,][]{2015MNRAS.447.1059K}, the 20~\kms{} cloud may have recently passed or will soon pass through the pericenter of an elliptical orbit, which results in a strong tidal effect. This tidal compression would drive turbulence and deform the cloud, during which transient substructures such as unbound over-densities may emerge. These substructures are not the usually bound dense cores, but are sometimes referred to as `starless cores' and will disperse over a few dynamical timescales \citep{offner2022}.
\end{itemize}

In this sense, the upcoming ALMA CMZ Exploration Survey (ACES) results, which have systematically surveyed the CMZ in continuum emission and molecular lines in the 3~mm band at a resolution of $\sim$2$''$, will be critical to discerning evolutionary phases and star formation states of clouds in the CMZ and exploring the origin of the different fractions of bound vs.\ unbound cores.

\subsection{Comparing the CMFs and the IMF in the CMZ}
\label{discussion-2}
In this study, we have updated the temperatures and therefore the masses of 19, 137, and 97 cores in cloud~e, Sgr~C, and the 20~\kms{} cloud, respectively, by fitting the \fmh{} or \chcn{} lines. \autoref{fig:moment_0} shows that the cores with \fmh{} or \chcn{} line emission are usually bright in the 1.3~mm continuum emission. In \citet{Lu2020ApJL}, when a universal dust temperature of 20~K was assumed, these cores were estimated to be massive and populate the high-mass ends of the CMFs. With the updated temperatures, their masses become lower, and thus the high-mass ends of the CMFs become steeper and become consistent with the shape of the Salpeter-like IMF, as presented in \autoref{subsec:results_CMFs}.

Additionally, \autoref{fig:cmf-Mmin=lu} illustrates that when the minimum masses of the power-law fitting are fixed to the same values in \citet{Lu2020ApJL}, the power-law indices remain to be higher than those reported in \citet{Lu2020ApJL}. Meanwhile, to verify the reliability of the data, we test the CMFs composed only of cores with the best-fit temperatures from the \fmh{} or \chcn{} lines, as shown in \autoref{fig:cmf-noly-updated_cores}. For each of the clouds, the number of cores is too low for a robust statistical result. For the three clouds combined, we again found a Salpeter-like shape in the high-mass end.

Therefore, using the updated core temperatures, the three clouds present a Salpeter-like slope in the high-mass end of their CMFs ($\alpha\sim1.35$). This changes the conclusion of top-heavy CMFs in \citet{Lu2020ApJL}, and highlights the importance of obtaining reliable core temperatures when fitting the CMFs \citep{dellova2024}. 

The Salpeter-like CMFs observed in the three CMZ clouds aligns with several recent findings toward clouds in the Galactic disk \citep[e.g.,][]{cao2021,suarez2021,Pouteau2023}, where Salpeter-like CMFs are also detected. However, unlike the Galactic disk where the IMF is found to be universally consistent with the Salpeter shape, the young massive star clusters in the CMZ are suggested to present top-heavy IMFs \citep{Husmann2012,Hosek2019}, although it can also be explained by tidal stripping of lower-mass stars \citep{park2020}. 

There have been debates on the inconsistency between CMFs and IMFs observed toward clouds in the Galactic disk, where the IMFs have the Salpeter-like slope in the high-mass end yet the CMFs sometimes are top-heavy \citep[e.g.,][]{zhang2015,Motte2018,Sanhueza2019,cheng2024,Louvet2024}. These results question the direct mapping between the CMF and the IMF, and may suggest a picture of dynamic evolution of core masses through active gas accretion that gradually converges the shape of the CMF to that of the IMF \citep{Pouteau2023,cheng2024}. In this study, the inconsistency possibly persists but the shapes of the two mass functions are reversed: the IMFs may be top-heavy, while the CMFs are found to be Salpeter-like. The implication is similar to the studies toward the Galactic disk regions: gas accretion into the cores as well as fragmentation may continuously change the shape of the CMF (e.g, more active accretion toward more massive cores: \citealt{Bonnell1998,Bonnell2001}; more efficient Jeans fragmentation of denser cores: \citealt{Larson2005,Elmegreen2008}), which may eventually lead to a top-heavy IMF.

\subsection{Caveats and perspectives}
We raise several caveats of our study:

\begin{itemize}
 \item Core identification: It has been shown that the parameters used in astrodendro affect the outcome of core identification and therefore the shape of CMFs \citep{Lu2020ApJL}. The algorithm adopted for the core identification (e.g., dendrogram vs.\ getsf) also has an impact on the derived CMFs \citep{cheng2024}. 
 \item Decoupled gas and dust temperatures: In the Galactic Center environment, the gas temperature and dust temperature are not necessarily well coupled. For example, in shocked regions in the 20 \kms{} cloud, an enhanced gas temperature has been observed \citep{Lu2017}, yet shock heating and pressure-volume work may not efficiently heat dust on $>$2000~AU scales. If a core is not internally heated by protostars, it may be possible that the gas temperature is higher than the dust temperature because of shock heating. In \autoref{appendix C}, we carry out a simple test to compare the \fmh{} emission that is or is not associated with the cores as traced by the 1.3~mm continuum, and find that the \fmh{} compact sources spatially associated with the cores present lower temperatures that the other sample that are presumably heated by shocks. Therefore, the \fmh{} gas in the cores seems to less affected by shock heating, although the possibility cannot be completely excluded.
 \item Overestimation of the velocity dispersions: As discussed in \autoref{subsubsec:results_fitting}, outflows traced by \fmh{} and hot cores traced by \chcn{} exist in the three clouds, which broaden the observed velocity dispersion and lead to overestimation of the virial parameters. In \autoref{fig:virial_histogram}, the few cores with virial parameters greater than 10 may be heavily contaminated by outflows or hot cores. Note that several commonly used tracers of dense gas in cores, such as N$_2$H$^+$, may be subject to the high ionization in the CMZ and therefore not well correlated with dense cores \citep{santamaria2021}.
 A careful selection of appropriate molecular lines that are correlated with dense gas in the cores and are less effected by outflows should be carried out, which will be the subject of next papers of the series.
 \item Incomplete temperature measurements in the cores: In \autoref{subsubsec:results_fitting}, we have updated the temperatures of a subset of the cores using the \fmh{} and \chcn{} lines, which results in a steeper slope for the high-mass ends of the CMFs. The remaining cores do not present sufficient S/N ratios in the \fmh{} or \chcn{} lines, yet many of them are found to be bound in \autoref{subsubsec:results_virial} and therefore could be already forming stars and be heated internally. Their temperatures, as a result, could be higher than the assumed 20~K, and then the core masses could be overestimated, although the temperatures cannot be too high (otherwise molecular transitions such as \fmh{} and \chcn{} should have been excited). The impact on the shape of the CMFs is an issue to be explored, should we obtain multiple-band continuum data in the future to construct dust spectral energy distributions and more robustly measure dust temperatures in the cores.
\end{itemize}

\section{Conclusions} 
\label{conclusions}
In this study, we investigate gas dynamics and CMFs of dense cores within three massive clouds in the CMZ using our ALMA 1.3~mm continuum and spectral line observations at 2000-AU resolution. We estimate temperatures and velocity dispersions using the \fmh{} or \chcn{} lines for a total of 253 cores in the cloud~e (19 out of 89), Sgr~C (137 out of 274), and the 20~\kms{} cloud (97 out of 471). Then we evaluate their virial equilibrium taking external pressure into consideration, and revisit their CMFs using the newly derived core temperatures as well as excluding unbound sources. Our main findings are:

\begin{itemize}

\item The high Mach numbers in the cores ($\mathcal{M}\sim$ 3--4) suggest the existence of supersonic non-thermal motions down to small spatial scales of 2000~AU.

\item In the extreme environment of the CMZ, the contribution of external pressure is critical for the virial equilibrium of the cores, without which most of the cores would be unbound. This differs from the situation of the cores in the Galactic disk where gas is already bound even without considering external pressure.

\item Compared with previous work where the core temperature is assumed to be 20~K universally \citep{Lu2020ApJL}, we update temperatures of the cores by fitting the \fmh{} and \chcn{} lines, and find that the high-mass ends of the CMFs of the three clouds exhibit a Salpeter-like slope. Given that the IMF in young massive star clusters in the CMZ may be top-heavy, this finding may suggest a picture of dynamic evolution of the CMF, including gas accretion and fragmentation, that eventually converges to the IMF.
\end{itemize}

\section*{Acknowledgments}
\begin{acknowledgments}
X.L.\ acknowledges support from the National Key R\&D Program of China (No.\ 2022YFA1603101), the Strategic Priority Research Program of the Chinese Academy of Sciences (CAS) Grant No.\ XDB0800300, the National Natural Science Foundation of China (NSFC) through grant Nos.\ 12273090 and 12322305, the Natural Science Foundation of Shanghai (No.\ 23ZR1482100), and the CAS ``Light of West China'' Program No.\ xbzg-zdsys-202212. A.G.\ acknowledges support from the National Science Foundation (NSF) under grants AAG 2008101 and 2206511 and CAREER 2142300. Y.C.\ was partially supported  by a Grant-in-Aid for Scientific Research (KAKENHI number JP24K17103) of the JSPS\@. H.B.L.\ is supported by the National Science and Technology Council (NSTC) of Taiwan (Grant Nos.\ 111-2112-M-110-022-MY3, 113-2112-M-110-022-MY3). Q.Z.\ acknowledges support from the NSF under Award No.\ 2206512. C.B.\ gratefully acknowledges  funding from the NSF under Award Nos.\ 2108938, 2206510, and CAREER 2145689, as well as from the National Aeronautics and Space Administration through the Astrophysics Data Analysis Program under Award ``3-D MC: Mapping Circumnuclear Molecular Clouds from X-ray to Radio,'' Grant No.\ 80NSSC22K1125. This paper makes use of the following ALMA data: ADS/JAO.ALMA\#2016.1.00243.S\@. ALMA is a partnership of ESO (representing its member states), NSF (USA), and NINS (Japan), together with NRC (Canada), MOST and ASIAA (Taiwan), and KASI (Republic of Korea), in cooperation with the Republic of Chile. The Joint ALMA Observatory is operated by ESO, AUI/NRAO, and NAOJ.\ This research has made use of NASA's Astrophysics Data System.
\end{acknowledgments}

\vspace{5mm}
\facility{ALMA}

\software{CASA \citep{McMullin2007,CASATeam2022}, APLpy \citep{Robitaille2012}, Astropy \citep{AstropyCollaboration2013,Astropy2018,Astropy2022}, XCLASS \citep{Moller2017}.}

\begin{deluxetable*}{@{\extracolsep{-7pt}}ccccccccccccccc}
\tabletypesize{\tiny}
\tablenum{1}
\tablewidth{\textwidth}
\tablecaption{Parameters of the cores and their \fmh{} and \chcn{} lines.\label{tab:three cloud only 1o}}
\tablehead{
\colhead{Core ID} &  \colhead{RA and Dec} & \colhead{Flux} & \colhead{$R$} & \colhead{$T_{\rm H_2CO}$} & \colhead{$T_{\rm H_2CO}$} & \colhead{$T_{\rm CH_3CN}$} & \colhead{$T$} & \colhead{$\sigma_{\rm H_2CO}$} & \colhead{$\sigma_{\rm H_2CO}$} & \colhead{$\sigma_{\rm CH_3CN}$} & \colhead{$\sigma$} & \colhead{$M_{\rm core}$} & \colhead{$\mathcal{M}$} &  \colhead{$\alpha_{\rm vir,p}$} \\
\colhead{} &  \colhead{} &  \colhead{density} & \colhead{} & \colhead{LTE} & \colhead{non-LTE} & \colhead{LTE} & \colhead{} & \colhead{LTE} &  \colhead{non-LTE} & \colhead{LTE} & \colhead{} & \colhead{} & \colhead{} \\
\colhead{} & \colhead{(J2000)}& \colhead{(mJy)} & \colhead{(AU)} & \colhead{(K)} & \colhead{(K)} & \colhead{(K)} & \colhead{(K)} & \colhead{(\kms{})} & \colhead{(\kms{})}& \colhead{(\kms{})}& \colhead{(\kms{})}& \colhead{(\msol{})}& \colhead{} & \colhead{}
}
\colnumbers
\startdata
cloud~e &  &  &   &  &  &  &  &  & & & & & & \\
1 & 17:46:47.19, $-$28:32:16.93 & 0.46 & 1200 & - & - & - & - & - & - & - & - & 0.69 & - & - \\ 
2 & 17:46:46.98, $-$28:32:14.43 & 0.48 & 1320 & - & - & - & - & - & - & - & - & 0.68 & - & - \\ 
3 & 17:46:48.39, $-$28:32:10.31 & 4.07 & 2250 & - & - & - & - & - & - & - & - & 6.12 & - & - \\ 
4 & 17:46:47.05, $-$28:32:09.91 & 0.72 & 1500 & 138.3$\pm$49.7 & 83$\pm$21 & - & 138.3$\pm$49.7 & 3.86$\pm$0.32 & 4.32$\pm$0.63 & - & 3.86$\pm$0.32 & 0.12 & 5.50$\pm$0.55 & 15.01 \\ 
5 & 17:46:47.07, $-$28:32:08.93 & 21.53 & 3950 & 69.2$\pm$14.7 & 73$\pm$0.5 & 94.4$\pm$9.7 & 94.4$\pm$9.7 & 2.73$\pm$0.25 & 1.36$\pm$0.24 & 3.02$\pm$0.19 & 3.02$\pm$0.19 & 3.93 & 5.21$\pm$0.29 & 4.76 \\ 
6 & 17:46:47.01, $-$28:32:08.88 & 0.25 & 1150 & - & - & - & - & - & - & - & - & 0.79 & - & - \\ 
7 & 17:46:46.93, $-$28:32:08.57 & 0.2 & 1030 & 95.4$\pm$34.6 & 70$\pm$5.5 & - & 95.4$\pm$34.6 & 1.34$\pm$0.19 & 1.34$\pm$0.42 & - & 1.34$\pm$0.19 & 0.05 & 2.09$\pm$0.38 & 2.31 \\ 
8 & 17:46:47.07, $-$28:32:07.26 & 152.95 & 8160 & - & - & 237.9$\pm$16.8 & 237.9$\pm$16.8 & - & - & 2.53$\pm$0.1 & 2.53$\pm$0.1 & 15.29 & 2.70$\pm$0.12 & 2.97 \\ 
9 & 17:46:46.95, $-$28:32:08.02 & 1.77 & 1820 & - & - & - & - & - & - & - & - & 2.53 & - & - \\ 
10 & 17:46:46.91, $-$28:32:07.58 & 1.1 & 1600 & - & - & - & - & - & - & - & - & 1.72 & - & - \\ 
Sgr~C &  &  &   &  &  &  &  &  & & & & & & \\
1 & 17:44:41.38, $-$29:28:29.90 & 19.33 & 4620 & 125.1$\pm$20.3 & 140$\pm$7 & - & 125.1$\pm$20.3 & 2.34$\pm$0.1 & 2.18$\pm$0.32 & - & 2.34$\pm$0.1 & 4.79 & 3.43$\pm$0.18 & 0.54 \\ 
2 & 17:44:41.32, $-$29:28:30.03 & 0.34 & 1360 & 88$\pm$28 & 101$\pm$63 & - & <89.1 & 1.56$\pm$0.2 & 1.27$\pm$0.37 & - & 1.56$\pm$0.2 & <0.09 & 2.61$\pm$0.41 & 0.26 \\ 
3 & 17:44:40.81, $-$29:28:27.70 & 0.37 & 1290 & - & - & - & - & - & - & - & - & 0.84 & - & - \\ 
4 & 17:44:41.21, $-$29:28:26.84 & 1.1 & 2330 & 113.5$\pm$42.8 & 125$\pm$16 & - & 113.5$\pm$42.8 & 1.56$\pm$0.18 & 1.43$\pm$0.53 & - & 1.56$\pm$0.18 & 0.31 & 2.30$\pm$0.33 & 0.13 \\ 
5 & 17:44:41.20, $-$29:28:23.94 & 0.61 & 1680 & 43.8$\pm$7.4 & 63$\pm$39 & - & 43.5$\pm$7.3 & 1.24$\pm$0.14 & 1.02$\pm$0.28 & - & 1.24$\pm$0.14 & 0.50 & 3.02$\pm$0.41 & 0.31 \\ 
6 & 17:44:40.76, $-$29:28:19.77 & 0.36 & 1370 & - & - & - & - & - & - & - & - & 0.77 & - & - \\ 
7 & 17:44:40.79, $-$29:28:19.50 & 1.15 & 1310 & 26.8$\pm$7.1 & 26$\pm$26 & - & <89.9 & 1.36$\pm$0.21 & 1.35$\pm$0.31 & - & 1.36$\pm$0.21 & >0.32 & 4.04$\pm$0.78 & 0.52 \\ 
8 & 17:44:40.78, $-$29:28:19.09 & 1.14 & 1310 & 73.7$\pm$31.2 & 130$\pm$12 & - & <99.5 & 1.7$\pm$0.33 & 1.53$\pm$0.63 & - & 1.7$\pm$0.33 & >0.28 & 3.15$\pm$0.72 & 0.35 \\ 
9 & 17:44:40.68, $-$29:28:19.20 & 0.43 & 1440 & - & - & - & - & - & - & - & - & 0.90 & - & - \\ 
10 & 17:44:40.73, $-$29:28:19.01 & 0.25 & 1150 & - & - & - & - & - & - & - & - & 0.54 & - & - \\ 
20~\kms{} cloud &  & &   &  &  &  &  &  & & & & & & \\
1 & 17:45:36.41, $-$29:06:30.01 & 0.4 & 1290 & - & - & - & - & - & - & - & - & 0.71 & - & - \\ 
2 & 17:45:36.53, $-$29:06:29.63 & 0.27 & 1040 & - & - & - & - & - & - & - & - & 0.36 & - & - \\ 
3 & 17:45:36.53, $-$29:06:29.38 & 0.33 & 1240 & - & - & - & - & - & - & - & - & 0.51 & - & - \\ 
4 & 17:45:36.23, $-$29:06:29.28 & 0.34 & 1330 & - & - & - & - & - & - & - & - & 0.55 & - & - \\ 
5 & 17:45:36.27, $-$29:06:28.57 & 0.25 & 1360 & - & - & - & - & - & - & - & - & 0.52 & - & - \\ 
6 & 17:45:36.64, $-$29:06:26.83 & 0.86 & 1720 & 101.5$\pm$49.7 & 125$\pm$61 & - & 101.5$\pm$49.7 & 2.39$\pm$0.37 & 2.99$\pm$0.96 & - & 2.39$\pm$0.37 & 0.21 & 3.90$\pm$0.70 & 6.12 \\ 
7 & 17:45:36.34, $-$29:06:26.57 & 0.26 & 1200 & - & - & - & - & - & - & - & - & 0.42 & - & - \\ 
8 & 17:45:36.70, $-$29:06:24.69 & 0.5 & 1790 & 61.2$\pm$20.4 & 69.1$\pm$3.1 & - & 61.2$\pm$20.4 & 3.43$\pm$0.51 & 3$\pm$1 & - & 3.43$\pm$0.51 & 0.21 & 7.34$\pm$1.18 & 11.18 \\ 
9 & 17:45:36.41, $-$29:06:24.27 & 0.33 & 1590 & - & - & - & - & - & - & - & - & 0.71 & - & - \\ 
10 & 17:45:36.31, $-$29:06:23.97 & 0.35 & 1290 & - & - & - & - & - & - & - & - & 0.46 & - & - \\ 
\enddata
\tablecomments{The table lists physical parameters of the first 10 dense cores from each of the three clouds complete lists are shown in Tables~\ref{tab:cloud e table}, \ref{tab:Sgr C cloud table}. Columns (2), (3), and (4) present the coordinates, 1.3~mm continnum flux density, and effective radius of the cores, respectively, as determined by astrodendro \citep{Lu2020ApJL}. Columns (5) to (7) show temperatures from the LTE fitting of \fmh{}, the non-LTE fitting of \fmh{}, and the LTE fitting of \chcn{} using EMANON. Column (8) lists the adopted temperature. Columns (9) to (11) correspond to the velocity dispersions from different approaches, and column (12) shows the adopted value. The masses, Mach numbers, and virial parameters of the cores are presented in columns (13), (14), and (15).}
\end{deluxetable*}

\clearpage

\appendix
\section{Fitting results of the \fmh{} and \chcn{} lines}
\label{appendix A}
\renewcommand\thefigure{A\arabic{figure}}
The left panel in \autoref{figA1} illustrates the fitting results of the \fmh{} lines using the LTE method to 233 cores, including 16, 127, and 90 cores in cloud~e, Sgr~C, and the 20~\kms{} cloud, respectively. The right panel shows the corresponding fitting results from the non-LTE method. Note that \autoref{figA1} includes \fmh{} fitting results of the cores with \chcn{} detections as well.

\begin{figure*}[htp!]
\newcounter{tempfigure} 
\setcounter{tempfigure}{\value{figure}} 
\setcounter{figure}{0} 
\renewcommand\thefigure{A\arabic{figure}} 
\centering
\includegraphics[height=6cm]{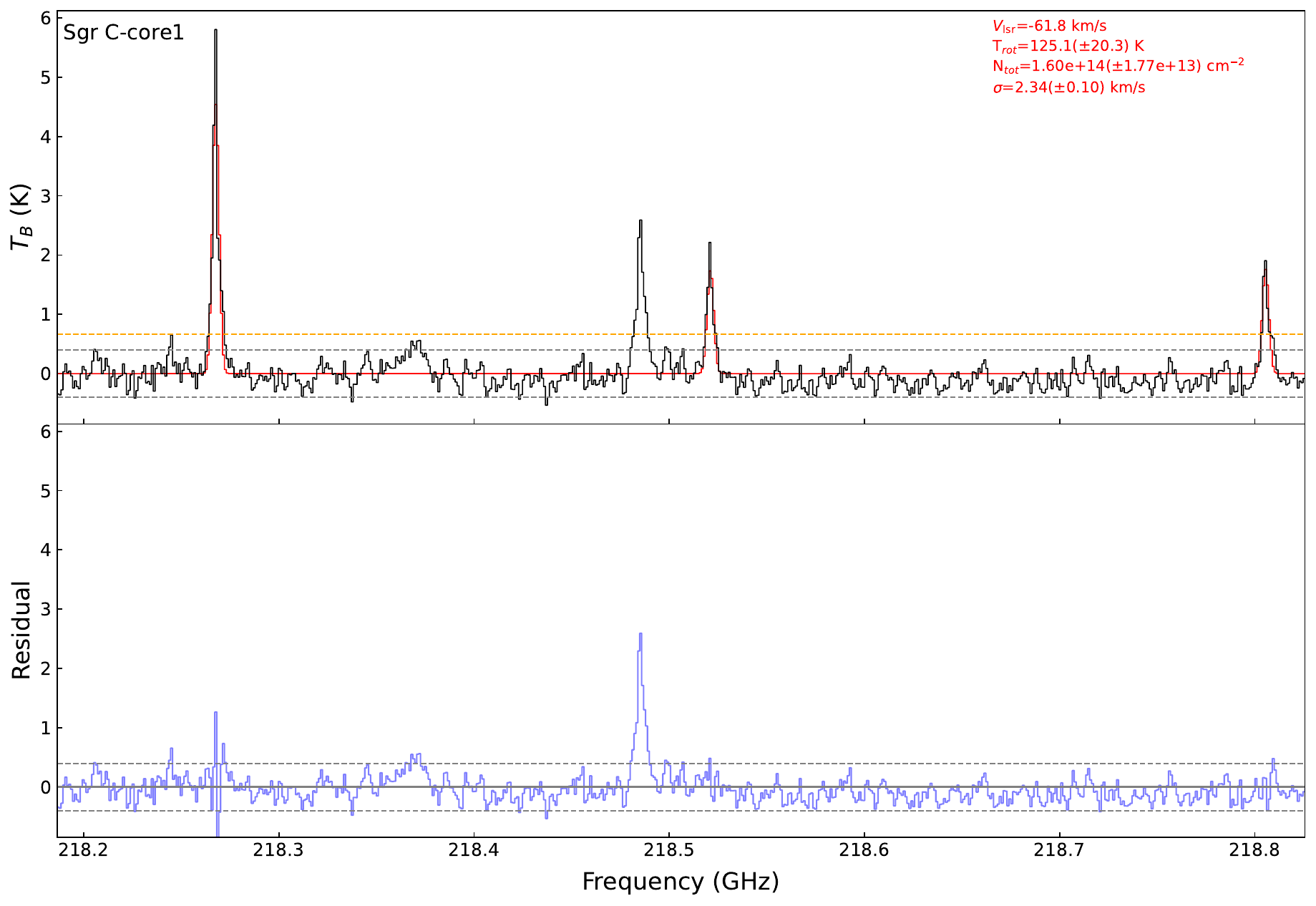}
\includegraphics[height=6cm]{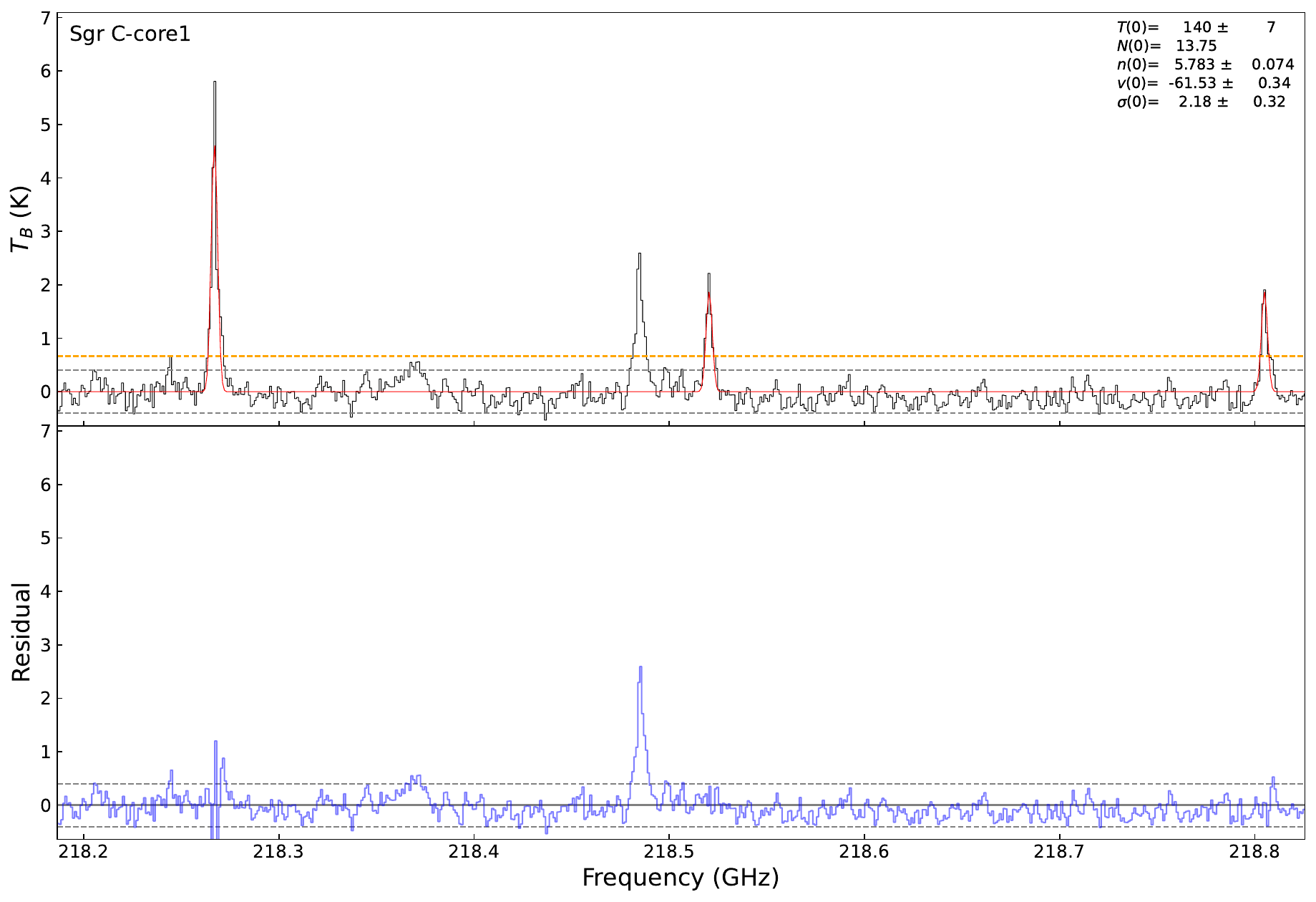}
\caption{The LTE and non-LTE fitting results towards one core are presented here. The complete figure set (466 images) is available in the online journal.}
\label{figA1}
\setcounter{figure}{\value{tempfigure}} 
\end{figure*}

\begin{figure*}[htp!]
\setcounter{tempfigure}{\value{figure}} 
\setcounter{figure}{1} 
\renewcommand\thefigure{A\arabic{figure}} 
\centering
\includegraphics[height=6cm]{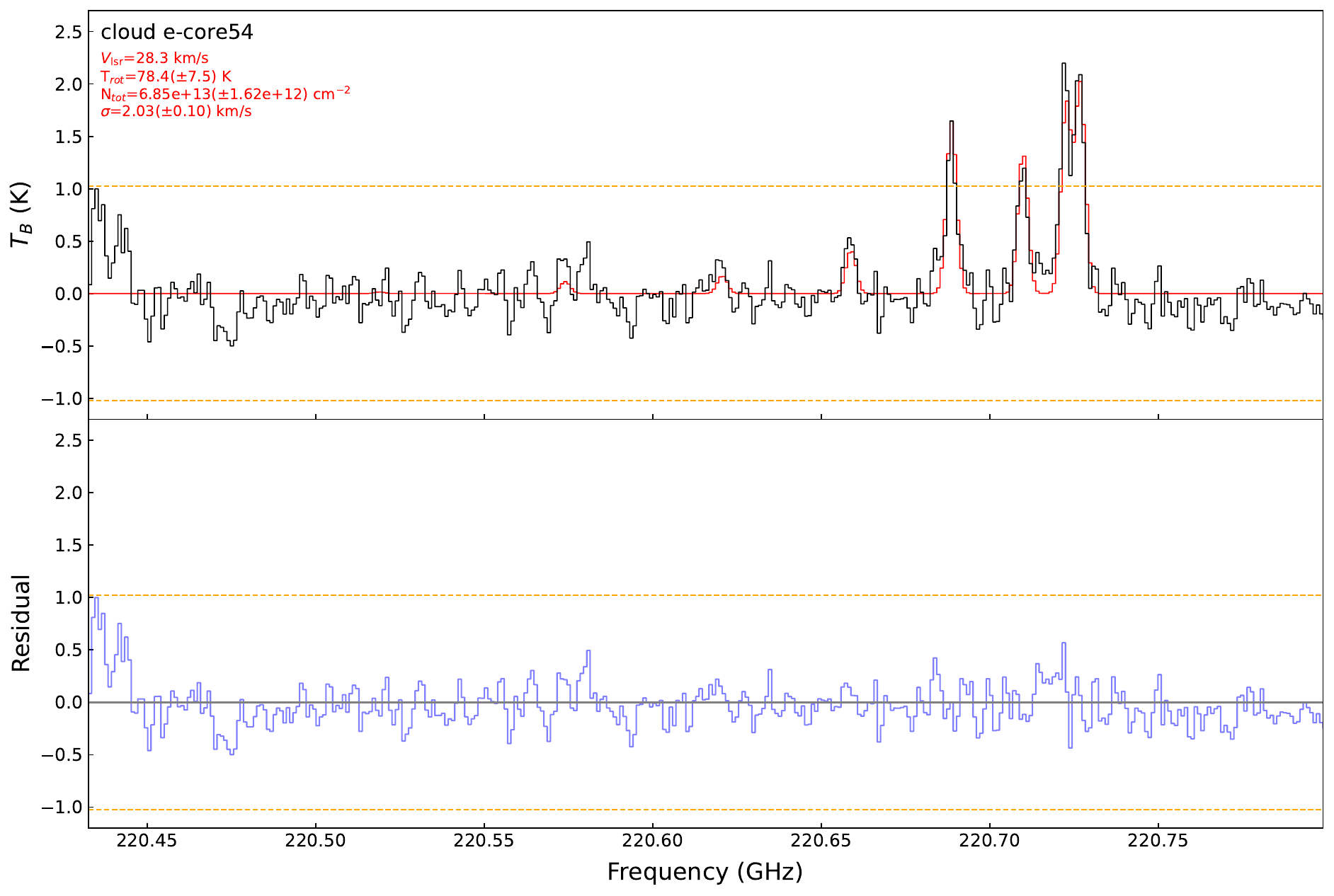}
\includegraphics[height=6cm]{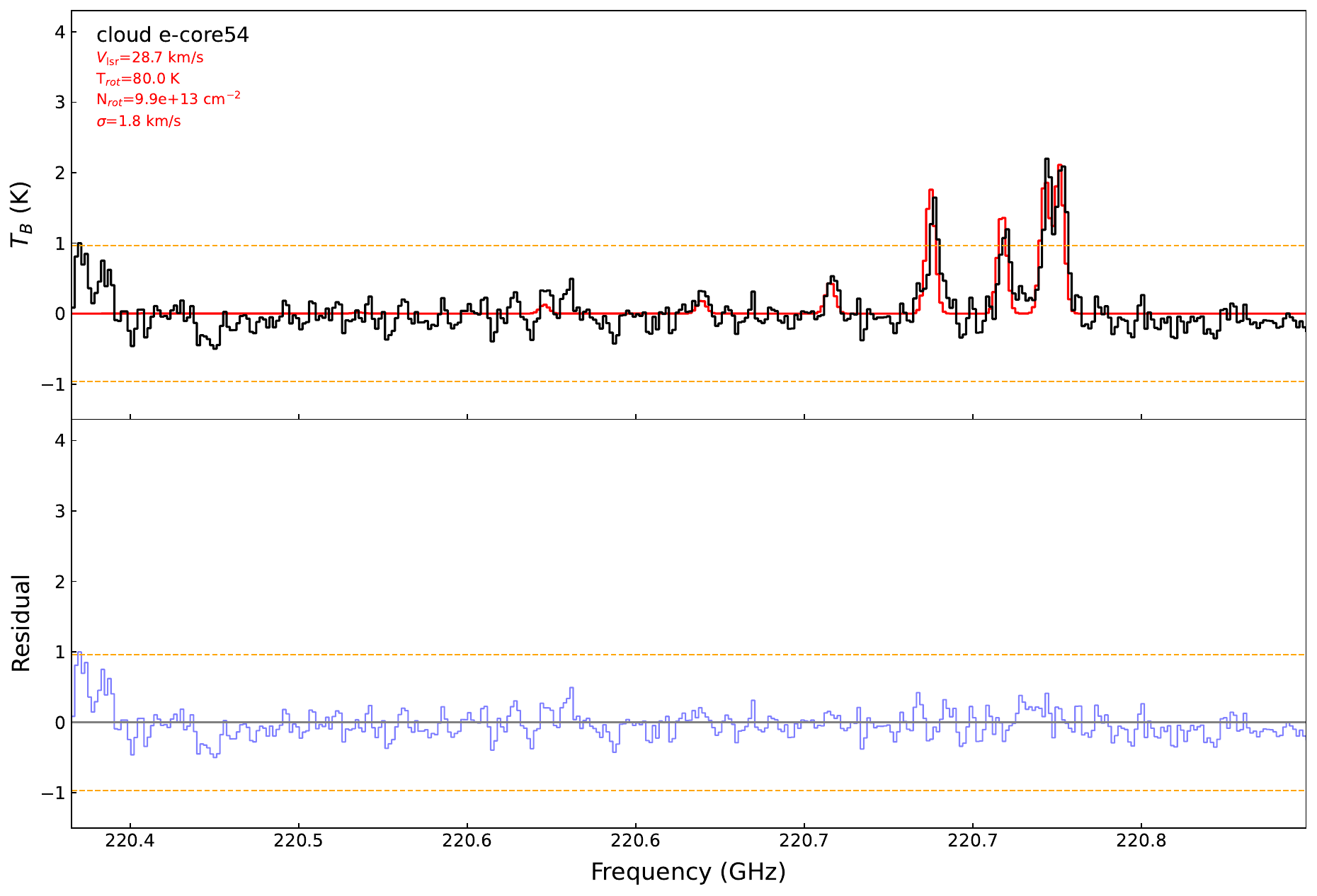}
\caption{The LTE fitting results using the EMANON code and XCLASS towards one core are presented here. The complete figure set (66 images) is available in the online journal.}
\label{figA2}
\setcounter{figure}{\value{tempfigure}} 
\end{figure*}

\autoref{figA2} depicts the LTE fittings of \chcn{} lines to 33 cores, including 4, 17, and 12 cores in cloud~e, Sgr~C, and the 20~\kms{} cloud, respectively. The left and right panels show the fitting results using the EMANON code and the XCLASS package, respectively.

\autoref{fig:T-linewidth} shows histograms of temperatures and velocity dispersions of the cores with \fmh{} spectra detected, comparing the LTE and non-LTE methods.

\autoref{fig:error-T-sigma} presents results from the LTE and non-LTE approaches employed for fitting the temperature and velocity dispersion using the \fmh{} lines. We observed a general consistency in the fitted temperatures and velocity dispersions between the two approaches. Note that under the assumption of LTE, we have assumed that the best-fit \fmh{} rotational temperature equals the kinetic temperature.

In \autoref{fig:error-T-sigma-CH3CN}, we compared temperatures derived using the two LTE approaches with \chcn{} lines. The results show that the temperatures are consistent within the fitting errors. 

\autoref{fig:T-error-H2COCH3CN} shows that the temperatures and velocity dispersions of the cores derived from \chcn{} are generally higher than those derived from \fmh{} (through forced fitting despite the self absorption). This may be because \chcn{} traces inner parts of a core where the gas is hotter and more turbulent due to stronger feedback from embedded protostars.

\begin{figure*}[htp!] 
\setcounter{tempfigure}{\value{figure}} 
\setcounter{figure}{2} 
\renewcommand\thefigure{A\arabic{figure}} 
\centering 
\includegraphics[height=0.22\textwidth]{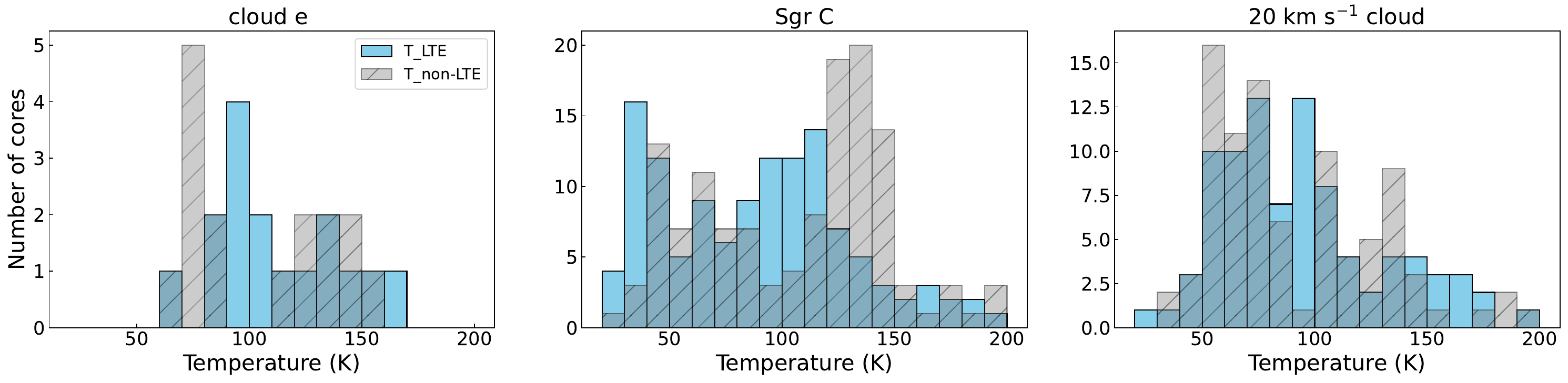}
\includegraphics[height=0.22\textwidth]{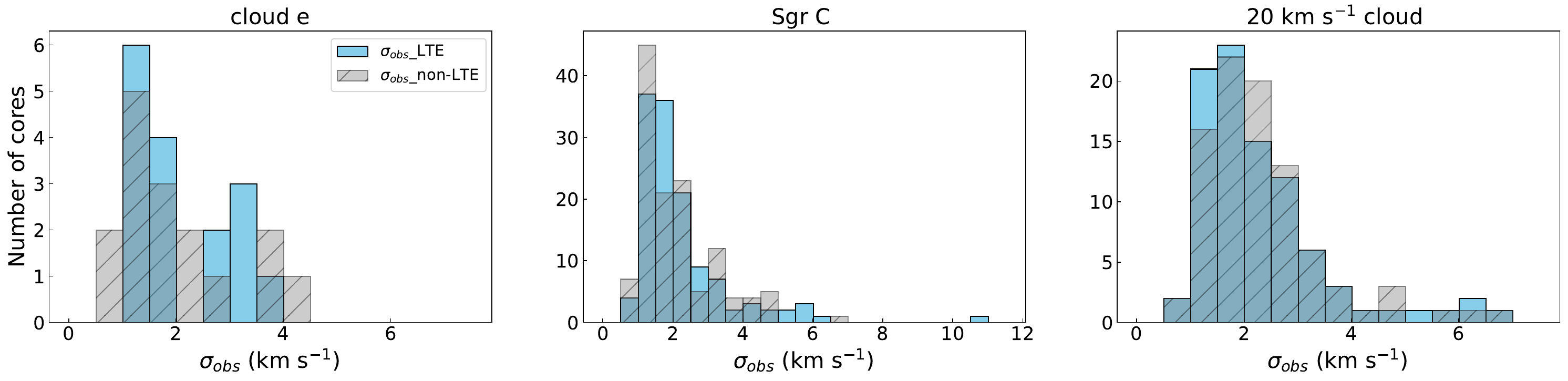}
\caption{Histograms of temperatures and velocity dispersions. The upper panel shows the temperatures, and the lower panels shows the velocity dispersions. The blue bars represent the LTE fitting results, while the striped grey bars represent the non-LTE fitting results.}
\label{fig:T-linewidth}
\setcounter{figure}{\value{tempfigure}} 
\end{figure*}

\begin{figure*}[htp!] 
\setcounter{tempfigure}{\value{figure}} 
\setcounter{figure}{3} 
\renewcommand\thefigure{A\arabic{figure}} 
\includegraphics[width=1\linewidth]{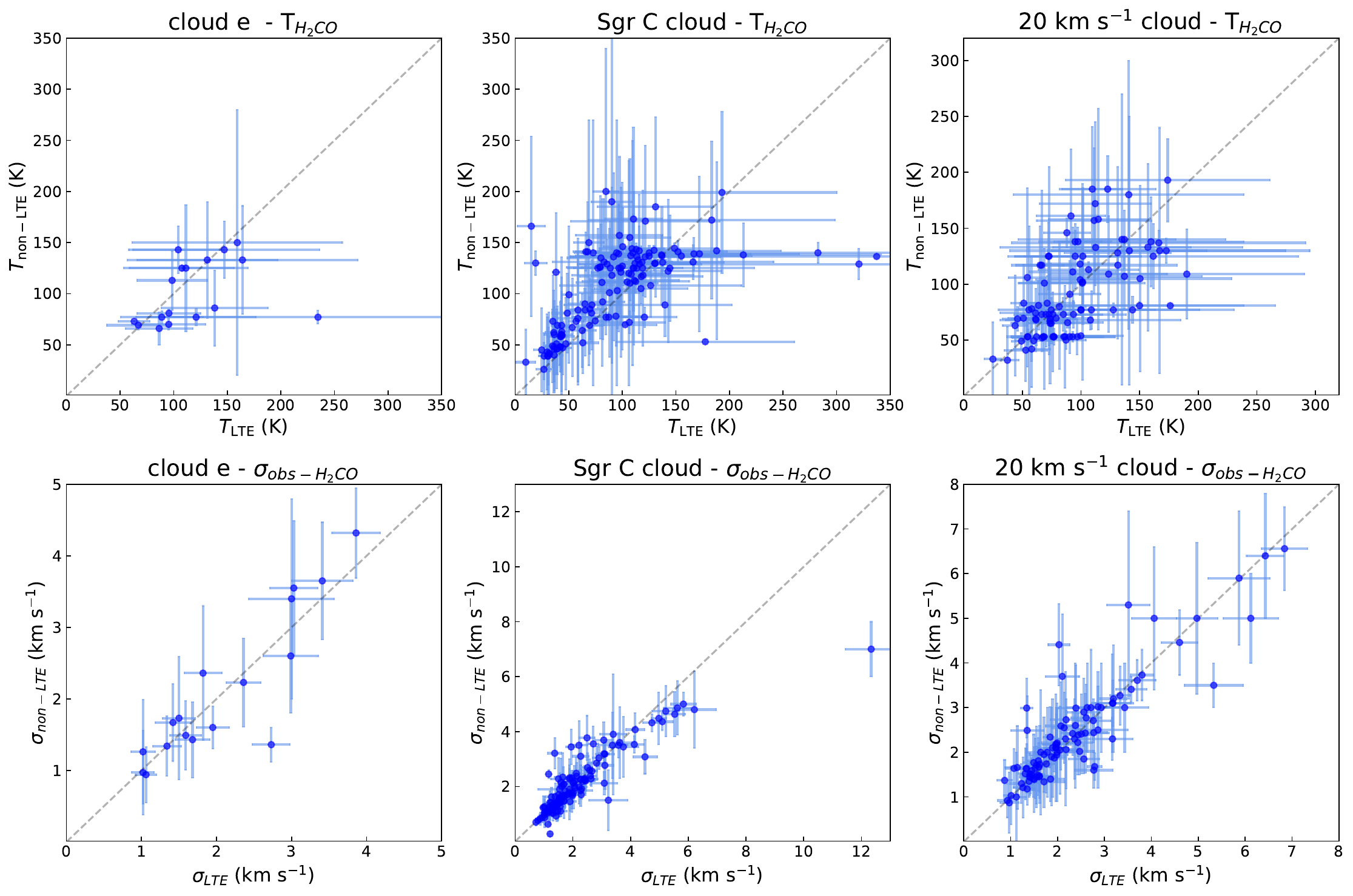}
 \caption{The upper panels show the correlation between the core temperatures in the three clouds derived from \fmh{} under LTE and non-LTE assumptions, while the lower panels depict the correlation for the velocity dispersions. The blue dots with error bars represent the best-fit values, and the dashed diagonal lines represent where the two fitting results are equal.}
\label{fig:error-T-sigma}
\setcounter{figure}{\value{tempfigure}} 
\end{figure*}

\begin{figure*}[htp!]  
\setcounter{tempfigure}{\value{figure}} 
\setcounter{figure}{4} 
\renewcommand\thefigure{A\arabic{figure}} 
\centering 
\includegraphics[width=1\linewidth]{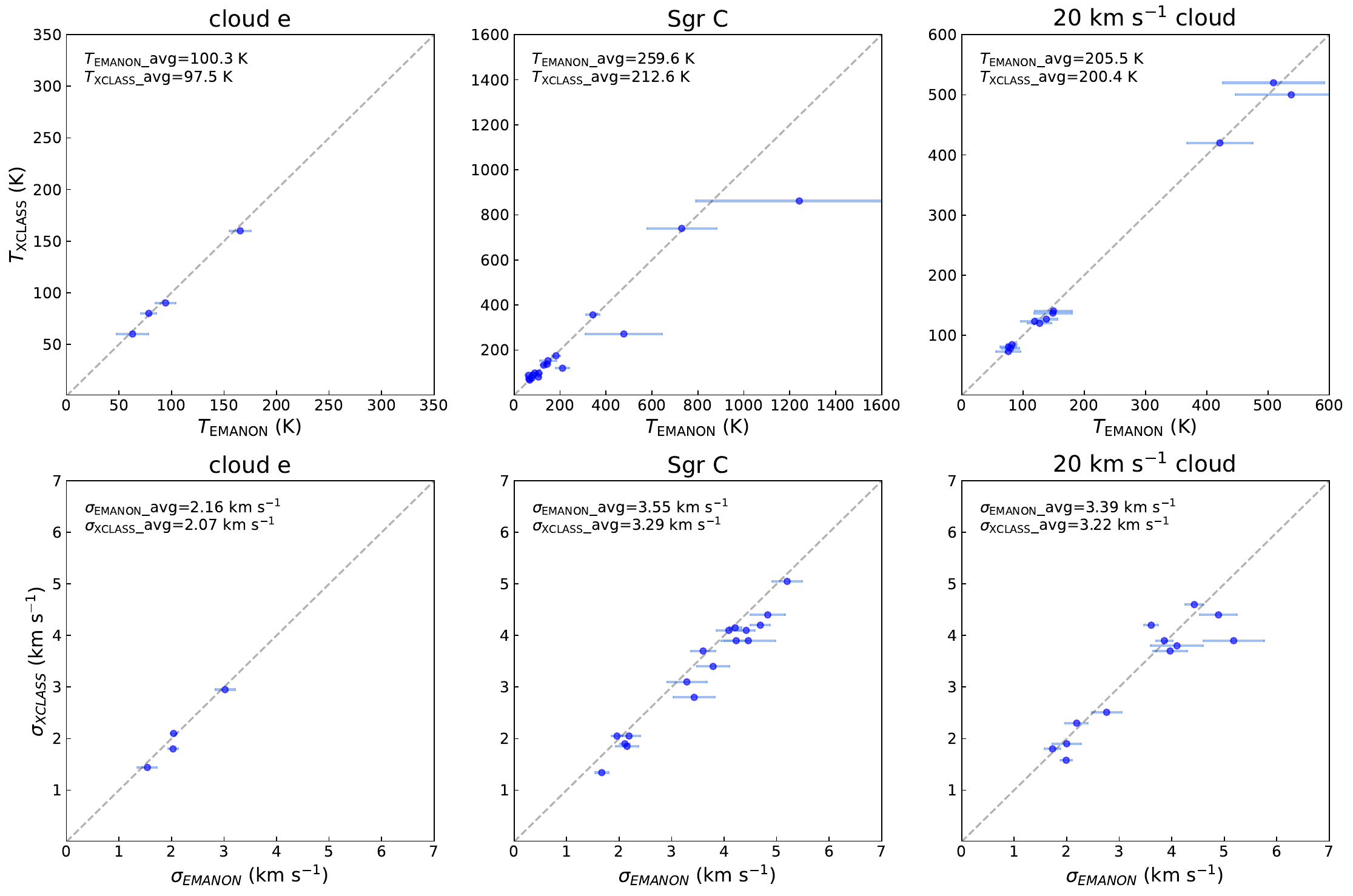}
 \caption{The upper panels show the correlation between the temperatures of the cores in the three clouds derived from \chcn{} under LTE conditions using the EMANON code vs.\ XCLASS, while the lower panels depict the correlation for the velocity dispersions. The symbols are the same as in \autoref{fig:error-T-sigma}, except that XCLASS fits do not output errors so vertical error bars are not plotted.}
\label{fig:error-T-sigma-CH3CN}
\setcounter{figure}{\value{tempfigure}} 
\end{figure*}

\begin{figure*}[htp!]
\setcounter{tempfigure}{\value{figure}} 
\setcounter{figure}{5} 
\renewcommand\thefigure{A\arabic{figure}} 
\centering
\includegraphics[height=7cm]{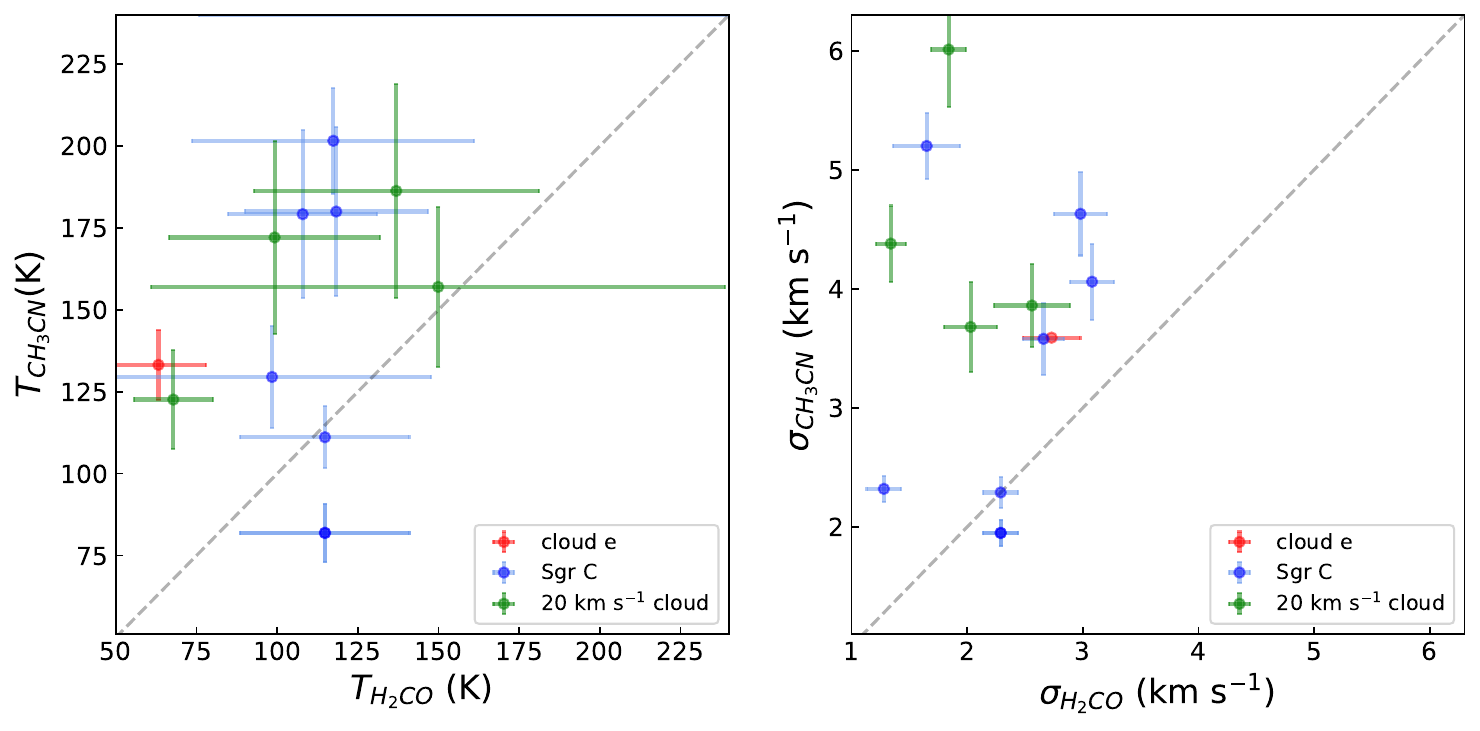}
\caption{The left panel shows the correlation between the core temperatures in the three clouds derived from \fmh{} and \chcn{} under LTE assumptions, while the right panel depicts the correlation for the velocity dispersions. The blue dots with error bars represent the best-fit values, and the dashed diagonal lines indicate where the fitting results are equal.}
\label{fig:T-error-H2COCH3CN}
\setcounter{figure}{\value{tempfigure}} 
\end{figure*}

\clearpage

\section{\fmh{} line ratio-temperature relation}\label{appendix B}
\renewcommand\thefigure{B\arabic{figure}}
In \autoref{subsubsec:results_fitting}, when only \fmh,$3_{0,3}$--$2_{0,2}$ line is detected (above the 5$\sigma$ level) in a core (i.e., scenario ii), we use the 3$\sigma$ value as the upper limit for the peak intensities of \fmh,$3_{2,2}$--$2_{2,1}$ and $3_{2,1}$--$2_{2,0}$ lines, and derive the upper limit for the line ratio between these two transitions and the $3_{0,3}$--$2_{0,2}$ transition. Then we collect all the cases where the $3_{0,3}$--$2_{0,2}$ line plus at least one of $3_{2,2}$--$2_{2,1}$ and $3_{2,1}$--$2_{2,0}$ lines are detected (i.e., scenario i), and fit their line ratios and temperature to derive a line ratio-temperature relation, as shown in \autoref{fig:lr_T}. The relation between the line ratio ($LR$) and the temperature ($T$) follows:
\begin{equation}
T = 67.9 \exp(0.8LR),
\end{equation}
which is then used to convert the line ratios to temperatures for scenario ii.

\begin{figure*}[htp!]
\setcounter{tempfigure}{\value{figure}} 
\setcounter{figure}{0} 
\renewcommand\thefigure{B\arabic{figure}} 
\centering
\includegraphics[height=6cm]{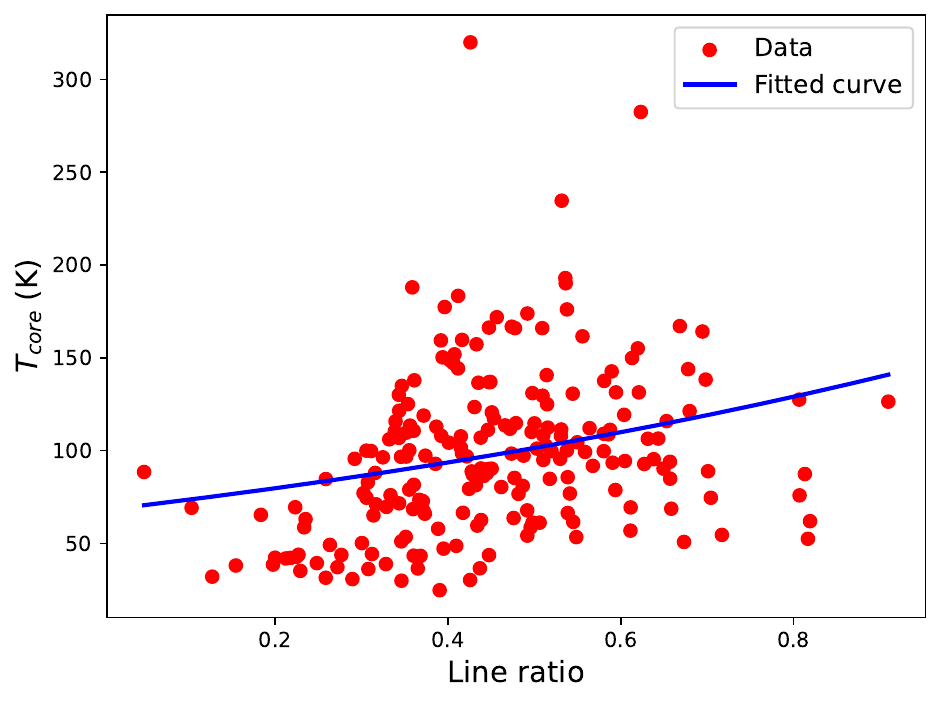}
\caption{\fmh{} line ratio-temperature relation map.}
\label{fig:lr_T}
\setcounter{figure}{\value{tempfigure}} 
\end{figure*}

\section{Correlation between the cores and \fmh{} gas}\label{appendix C}

\renewcommand\thefigure{C\arabic{figure}}
As discussed in \autoref{subsec:results_lineemission}, the \fmh{} line emission is not necessarily associated with the cores traced by the 1.3~mm continuum. To test how well the \fmh{} emission and the 1.3~mm continuum emission are correlated, we select the C1 clump in the 20~\kms cloud as an example, where apparent spatially extended \fmh{} emission is seen. We use astrodendro to identify 96 compact sources in the integrated intensity map of the \fmh{} $3_{0,3}$--$2_{0,2}$ line. Then we clarify them into those associated with the dense cores (`core-related') and those not (`core-unrelated'), which consist of 10 and 86 compact sources, respectively, and fit the \fmh{} lines. However, 67 out of these sources cannot be fit because their $3_{0,3}$--$2_{0,2}$ peak intensities are below 5$\sigma$. \autoref{fig:C1} shows best-fit temperatures and velocity dispersions of the compact sources where LTE fittings can be carried out. The core-related compact sources clearly present lower temperatures that core-unrelated ones.

We further run K-S tests on the temperatures and velocity dispersions of the two samples. As shown in \autoref{fig:C2}, the $p$ value of the K-S test on the temperatures is 0.068, suggesting a marginally significant difference between the core-related and core-unrelated compact sources. However, the difference in velocity dispersions is not statistically significant given the high $p$ value.

Therefore, the \fmh{} emission spatially not associated with the cores may indeed represent a hotter gas component than the dense gas in the cores. Meanwhile, whether \fmh{} detected toward the cores solely traces dense gas remains uncertain, although it may present statistically different temperatures than the core-unrelated component.

\begin{figure*}[htp!]
\setcounter{tempfigure}{\value{figure}} 
\setcounter{figure}{0} 
\renewcommand\thefigure{C\arabic{figure}} 
\centering
\includegraphics[height=6cm]{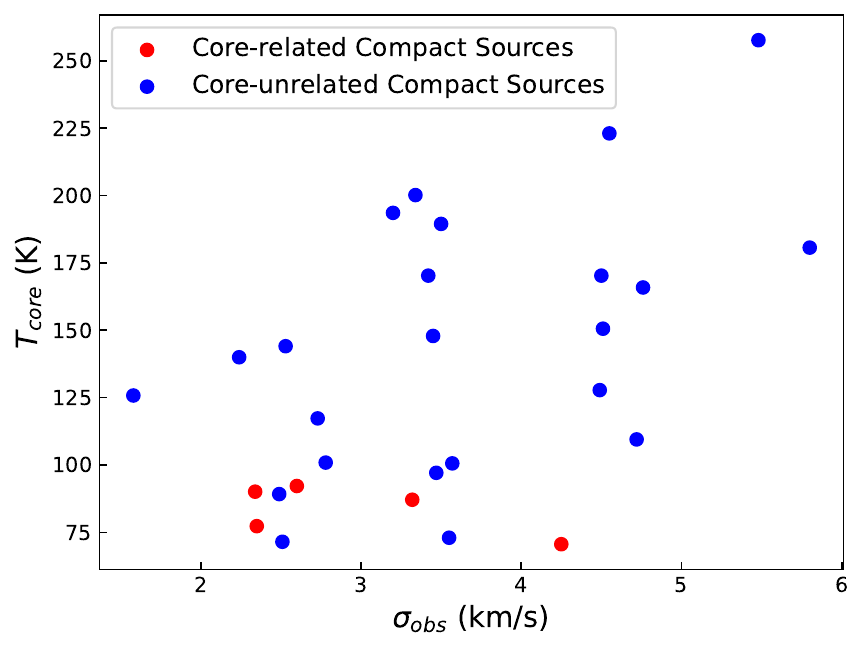}
\caption{Temperatures and velocity dispersions of the \fmh{} compact sources. The red points represent the core-related \fmh{} compact sources, while the blue points indicate the core-unrelated ones.}
\label{fig:C1}
\setcounter{figure}{\value{tempfigure}} 
\end{figure*}

\begin{figure*}[htp!]
\setcounter{tempfigure}{\value{figure}} 
\setcounter{figure}{1} 
\renewcommand\thefigure{C\arabic{figure}} 
\centering
\includegraphics[height=6cm]{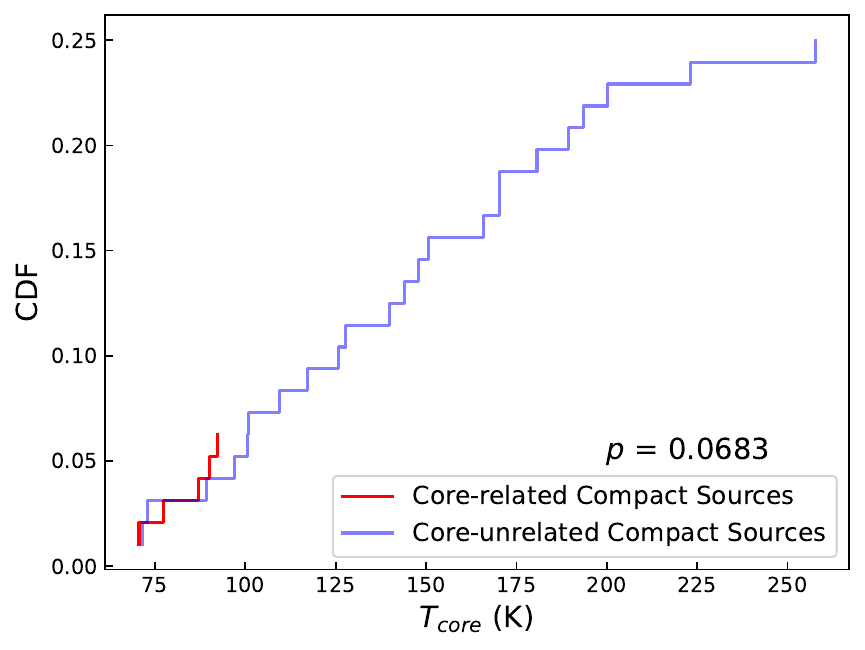}
\includegraphics[height=6cm]{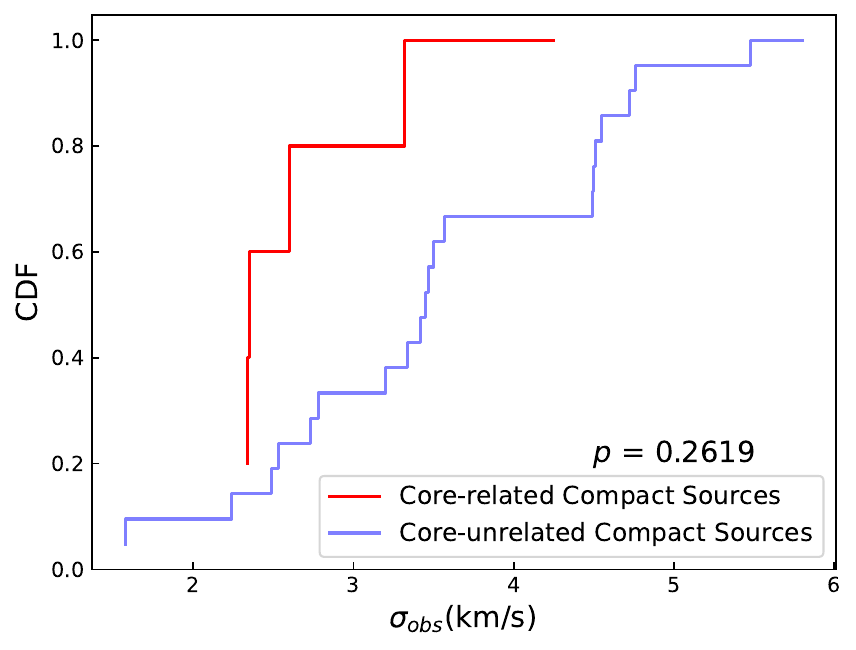}
\caption{The left panel shows the cumulative distribution of temperatures for both core-related and core-unrelated \fmh{} compact sources, while the right panel shows the cumulative distribution of velocity dispersions.}
\label{fig:C2}
\setcounter{figure}{\value{tempfigure}} 
\end{figure*}

\section{Tests on the Core Mass Functions}\label{appendix D}

We carry out a series of tests on the CMFs, most of which are about different assumptions on the dust temperature. In all cases, the power-law indices of the high-mass ends of the CMFs remain to be higher than those derived in \citet{Lu2020ApJL} and close to 1.35.

\autoref{fig:cmf-Mmin=lu} shows the CMFs for the three individual clouds and the three clouds combined, adopting the core temperatures derived in \autoref{subsec:results_Tcore} and fixing the minimum masses for the power-law fit to the same ones in \citet{Lu2020ApJL}. This demonstrates that the higher power-law indices in this work are not a result of different minimum masses in the fit.

\begin{figure*}[htp!]
\setcounter{tempfigure}{\value{figure}} 
\setcounter{figure}{0} 
\renewcommand\thefigure{D\arabic{figure}} 
\centering
\includegraphics[width=0.7\textwidth]{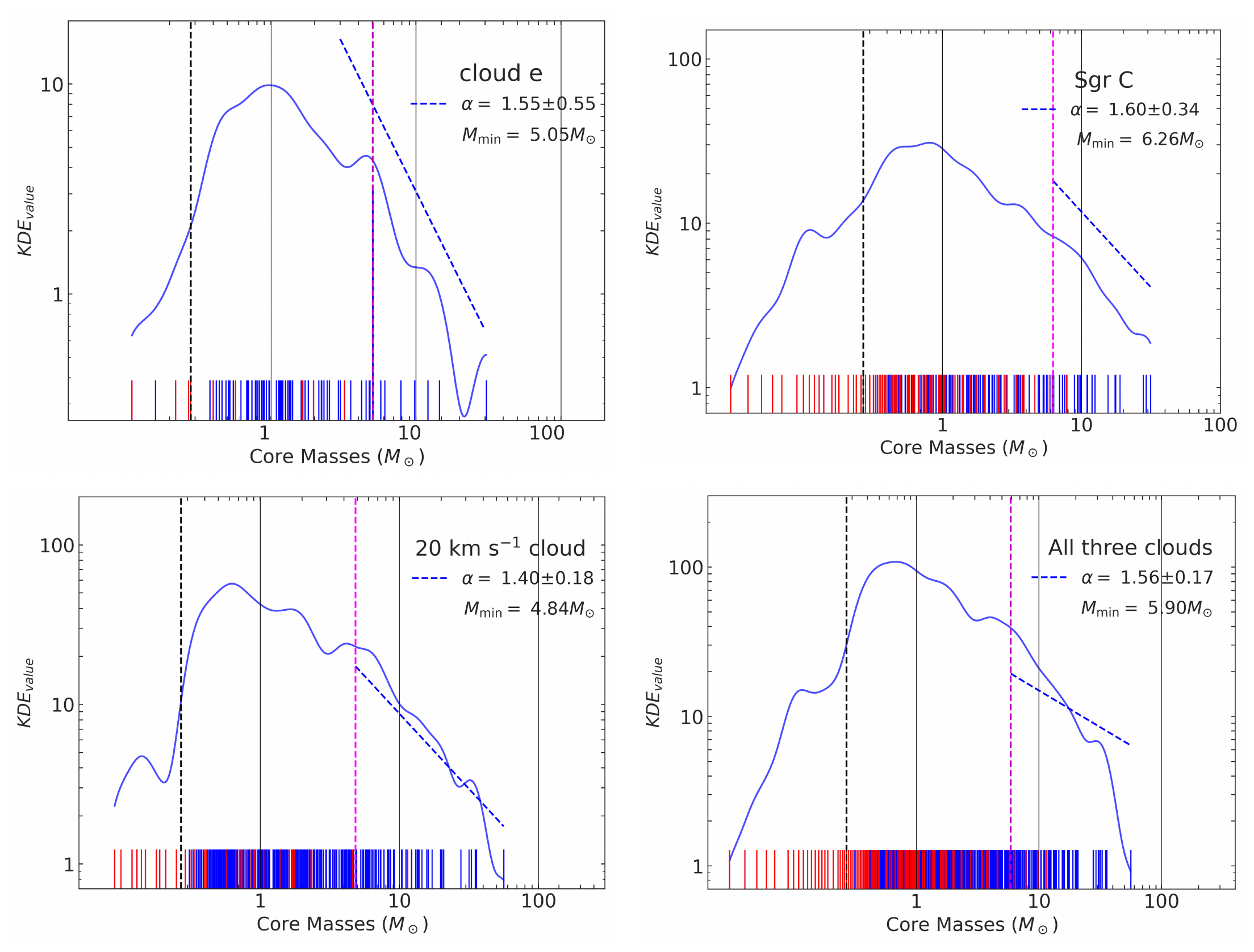}
\caption{The CMFs for the three individual clouds and the three clouds combined after fixing the minimum mass for the power-law fit. The symbols in the figures are the same as in \autoref{fig:cmf}.}
\label{fig:cmf-Mmin=lu}
\setcounter{figure}{\value{tempfigure}} 
\end{figure*}

\autoref{fig:cmf-T-assume-30K} shows the CMFs for the three individual clouds and the three clouds combined, where core temperatures derived in \autoref{subsec:results_Tcore} are used, and for cores without \fmh{} or \chcn{} detections, core temperatures of 30~K are adopted. Similarly, \autoref{fig:cmf-T-assume-50K} shows the case where core temperatures of 50~K are adopted when \fmh{} or \chcn{} lines are not detected. The power law indices are higher than those reported in \citet{Lu2020ApJL} and can be up to $\sim$2.

\begin{figure*}[htp!]
\setcounter{tempfigure}{\value{figure}} 
\setcounter{figure}{1} 
\renewcommand\thefigure{D\arabic{figure}} 
\centering
\includegraphics[width=0.7\textwidth]{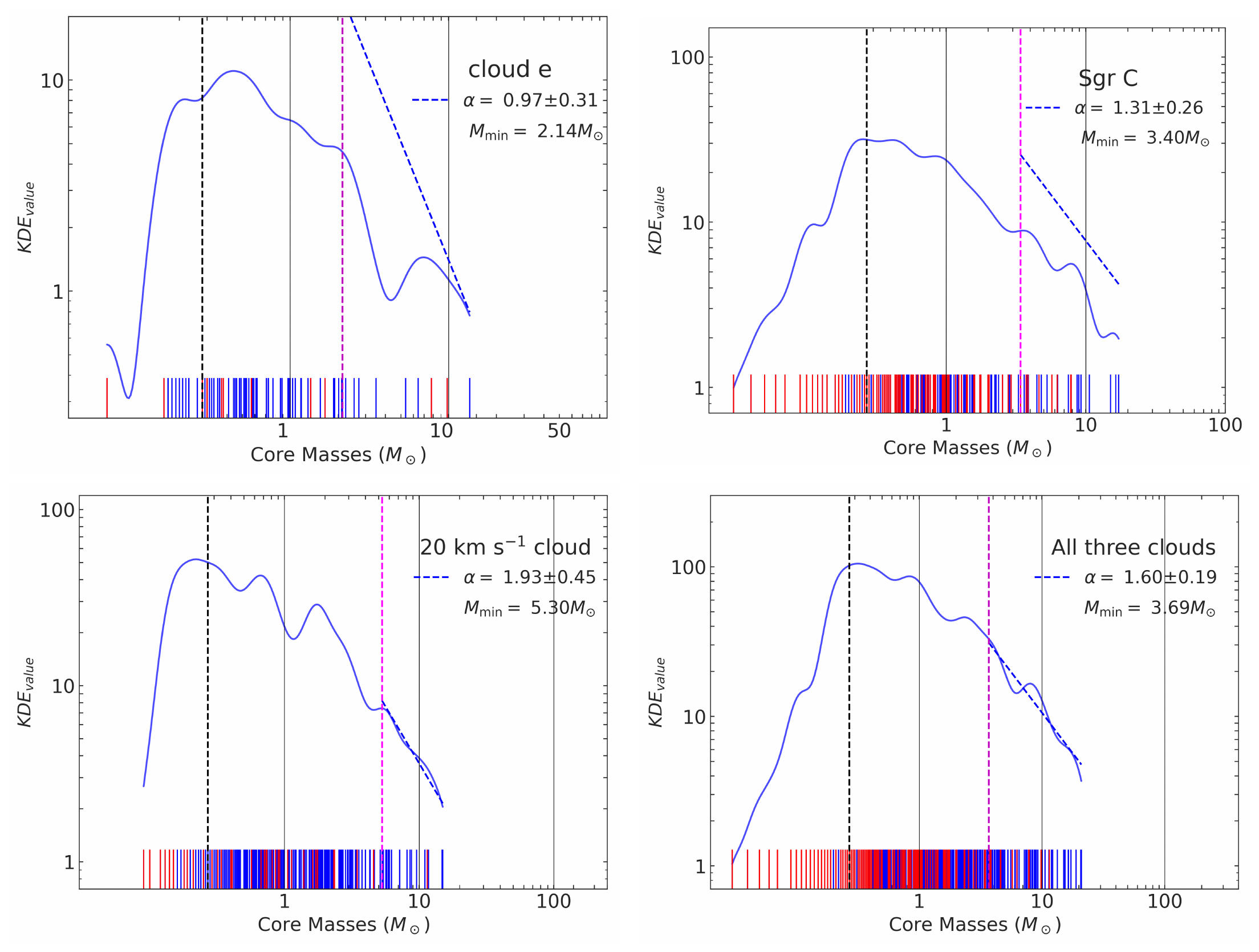}
\caption{The CMFs for the three individual clouds and the three clouds combined, with core temperatures of 30~K adopted for those without \fmh{} or \chcn{} detections. The symbols in the figures are the same as in \autoref{fig:cmf}.} 
\label{fig:cmf-T-assume-30K}
\setcounter{figure}{\value{tempfigure}} 
\end{figure*}

\begin{figure*}[htp!]
\setcounter{tempfigure}{\value{figure}} 
\setcounter{figure}{2} 
\renewcommand\thefigure{D\arabic{figure}} 
\centering
\includegraphics[width=0.7\textwidth]{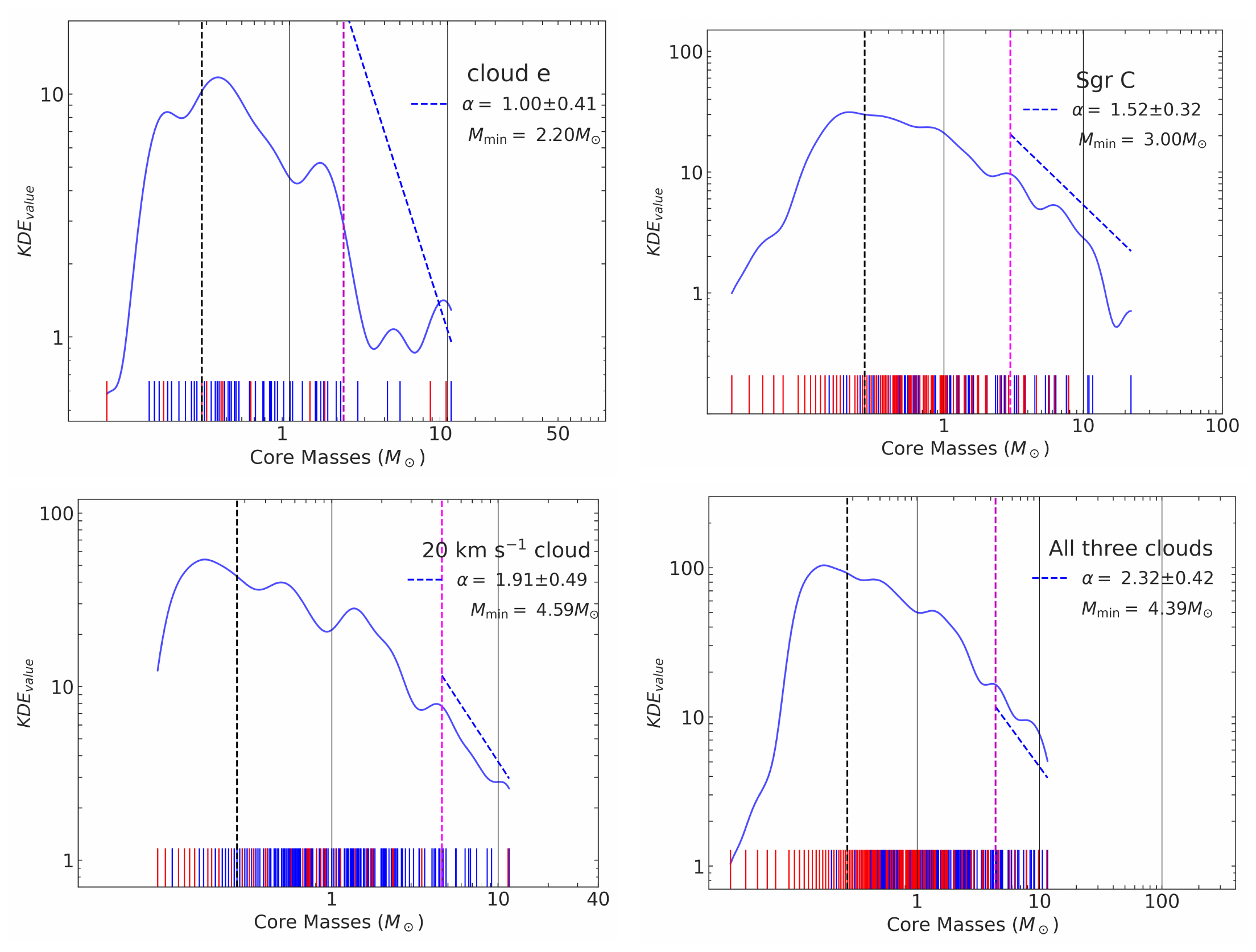}
\caption{The CMFs for the three individual clouds and the three clouds combined, with core temperatures of 50~K adopted for those without \fmh{} or \chcn{} detections. The symbols in the figures are the same as in \autoref{fig:cmf}.}
\label{fig:cmf-T-assume-50K}
\setcounter{figure}{\value{tempfigure}} 
\end{figure*}

\autoref{fig:cmf-noly-updated_cores} shows the CMFs for the three individual clouds and the three clouds combined, adopting the core temperatures derived in \autoref{subsec:results_Tcore}. Here we only include the core masses with updated temperatures, and perform the power-law fitting to the high-mass ends of the CMFs following the same approach as in \autoref{subsec:results_CMFs}. For the three clouds combined, one finds a Salpeter-like shape in the high-mass end of the CMF. We note that this test does not imply a new selection criteria that only considers warmer or hotter cores with \fmh{} and/or \chcn{} emission, but only illustrates that the CMF shape is not affected by excluding the colder cores in \autoref{fig:cmf}.

\begin{figure*}[htp!]
\setcounter{tempfigure}{\value{figure}} 
\setcounter{figure}{3} 
\renewcommand\thefigure{D\arabic{figure}}
\centering
\includegraphics[width=0.7\textwidth]{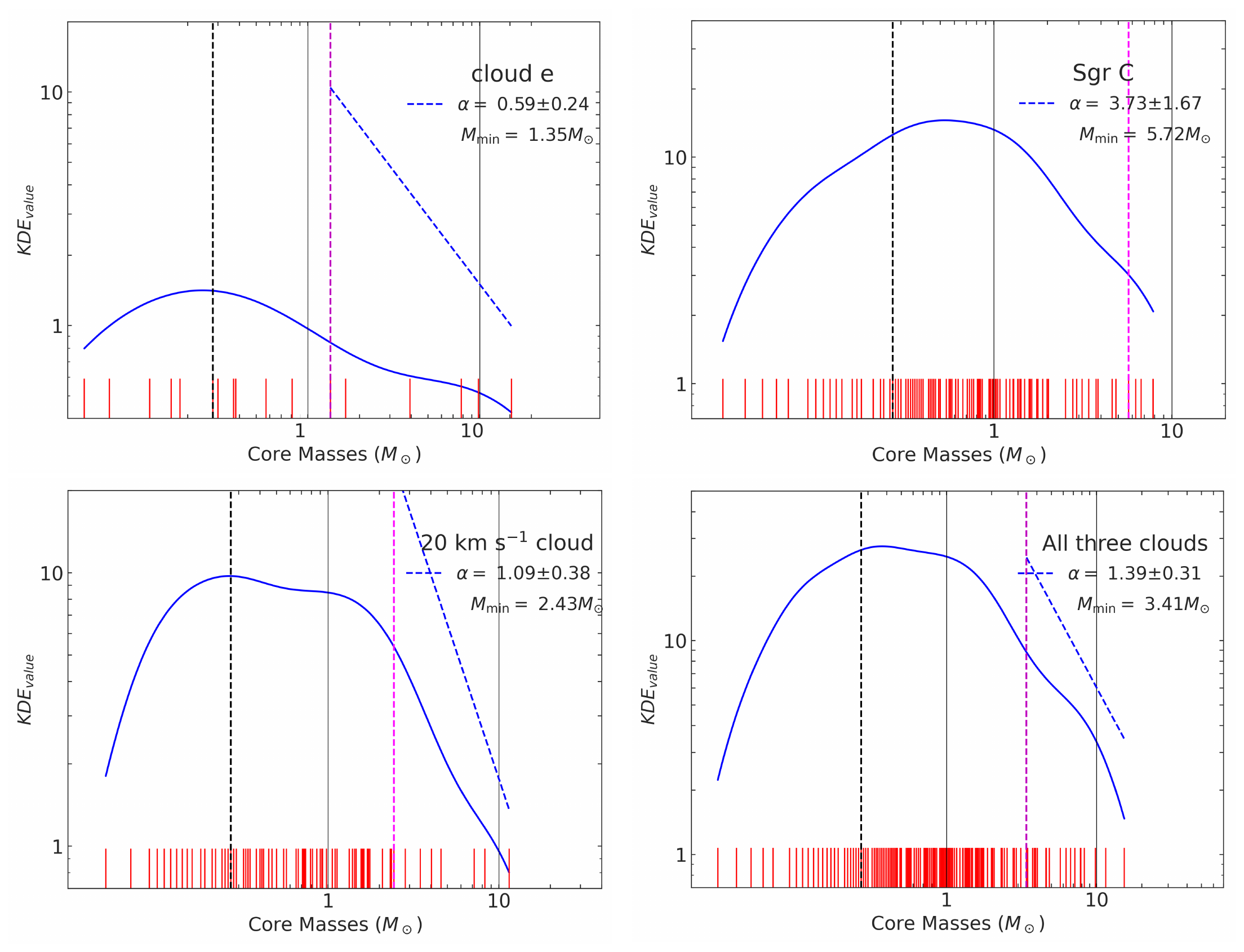}
\caption{The CMFs for the three individual clouds and the three clouds combined, with only the core masses derived from best-fit temperatures using \fmh{} or \chcn{} lines. The symbols in the figures are the same as in \autoref{fig:cmf}.}
\label{fig:cmf-noly-updated_cores}
\setcounter{figure}{\value{tempfigure}} 
\end{figure*}

We also test whether adopting the core temperatures derived from non-LTE fits of the \fmh{} lines has any effect on the slope of the CMFs. \autoref{fig:cmf-non-LTE} demonstrates that the results are consistent with those in \autoref{fig:cmf}.

\begin{figure*}[htp!]
\setcounter{tempfigure}{\value{figure}} 
\setcounter{figure}{4} 
\renewcommand\thefigure{D\arabic{figure}} 
\centering
\includegraphics[width=0.7\textwidth]{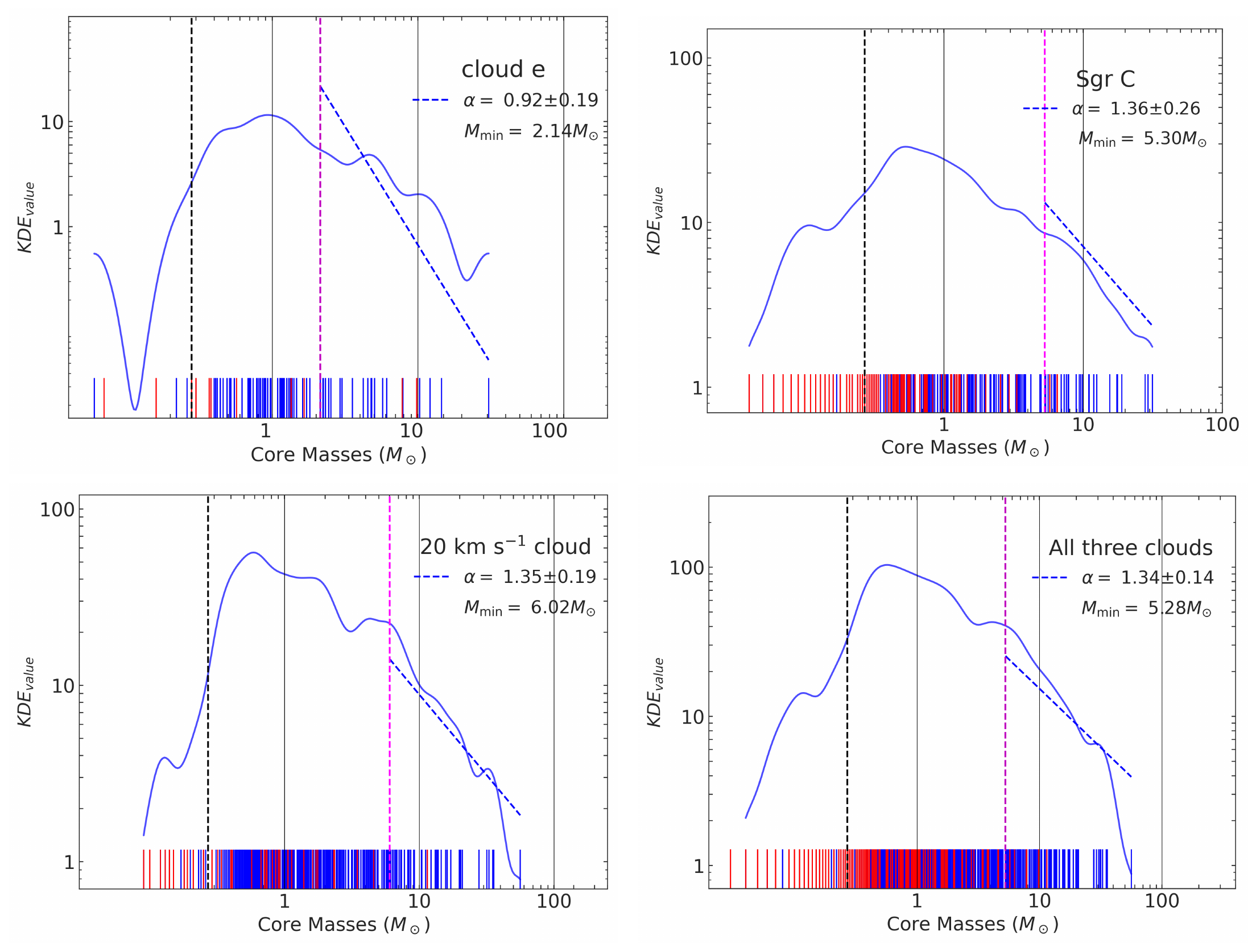}
\caption{The CMFs for the three individual clouds and the three clouds combined, with core temperatures derived from the non-LTE fitting of \fmh{} instead of from the LTE fitting, otherwise the core masses are identical to those in \autoref{fig:cmf}. The symbols in the figures are the same as in \autoref{fig:cmf}.}
\label{fig:cmf-non-LTE}
\setcounter{figure}{\value{tempfigure}} 
\end{figure*}

Lastly, we carry out a Monte Carlo experiment by varying the core temperatures within their corresponding uncertainty ranges, and regenerating the core masses and the CMFs 80000 times. As shown in \autoref{fig:monte-carlo}, the power law indices of the CMFs of the three clouds show Gaussian distributions around mean values of 1.2--1.4, consistent with being Salpeter-like.

\begin{figure*}[htp!]
\setcounter{tempfigure}{\value{figure}} 
\setcounter{figure}{5} 
\renewcommand\thefigure{D\arabic{figure}} 
\centering
\includegraphics[width=1\textwidth]{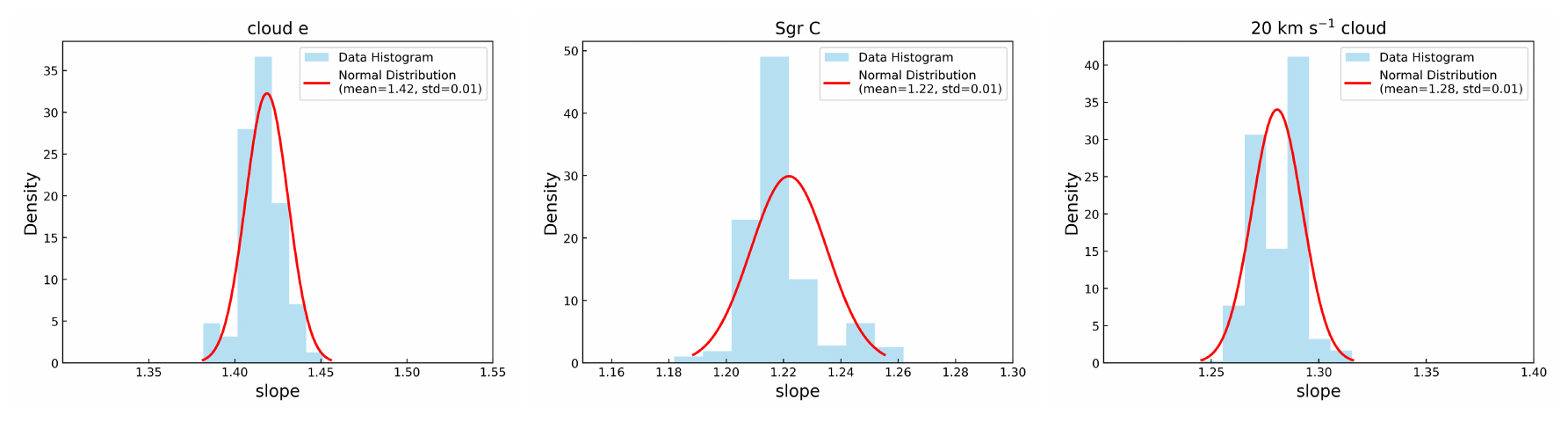}
\caption{Distribution of the slopes of the CMFs in the three molecular clouds from Monte Carlo tests. The blue bars represent the data points and the red curves represent the Gaussian fits. The mean and standard deviation of each cloud are shown in the upper right corner of each panel.}
\label{fig:monte-carlo} 
\setcounter{figure}{\value{tempfigure}} 
\end{figure*}

\clearpage

\section{Full Core catalogs}\label{appendix E}
The following three tables list properties of the cores in cloud~e, Sgr~C, and the 20~\kms{} cloud. The data columns are the same as those in \autoref{tab:three cloud only 1o}.

\startlongtable
\begin{deluxetable*}{@{\extracolsep{-8pt}}ccccccccccccccc}
\tabletypesize{\tiny}
\tablenum{E1}
\tablewidth{\textwidth}
\tablecaption{Properties of the cores in cloud e.\label{tab:cloud e table}}
\tablehead{
\colhead{Core ID} &  \colhead{RA and Dec} & \colhead{Flux} & \colhead{$R$} & \colhead{$T_{\rm H_2CO}$} & \colhead{$T_{\rm H_2CO}$} & \colhead{$T_{\rm CH_3CN}$} & \colhead{$T$} & \colhead{$\sigma_{\rm H_2CO}$} & \colhead{$\sigma_{\rm H_2CO}$} & \colhead{$\sigma_{\rm CH_3CN}$} & \colhead{$\sigma$} & \colhead{$M_{\rm core}$} & \colhead{$\mathcal{M}$} &  \colhead{$\alpha_{\rm vir,p}$} \\
\colhead{} &  \colhead{} &  \colhead{density} & \colhead{} & \colhead{LTE} & \colhead{non-LTE} & \colhead{LTE} & \colhead{} & \colhead{LTE} &  \colhead{non-LTE} & \colhead{LTE} & \colhead{} & \colhead{} & \colhead{} \\
\colhead{} & \colhead{(J2000)}& \colhead{(mJy)} & \colhead{(AU)} & \colhead{(K)} & \colhead{(K)} & \colhead{(K)} & \colhead{(K)} & \colhead{(\kms{})} & \colhead{(\kms{})}& \colhead{(\kms{})}& \colhead{(\kms{})}& \colhead{(\msol{})}& \colhead{} & \colhead{}
}
\colnumbers
\startdata
1 &17:46:47.19, -28:32:16.93 &0.46& 1200 &- &- &- &- &- &-&-&-&0.69&-&-\\
2 &17:46:46.98, -28:32:14.43 &0.48& 1320 &- &- &- &- &- &-&-&-&0.68&-&-\\
3 &17:46:48.39, -28:32:10.31 &4.07& 2250 &- &- &- &- &- &-&-&-&6.12&-&-\\
4 &17:46:47.05, -28:32:09.91 &0.72& 1500 &138.3±49.7 &83±21 &- &138.3±49.7 &3.86±0.32 &4.32±0.63&-&3.86±0.32&0.12&5.50±0.55&15.01\\
5 &17:46:47.07, -28:32:08.93 &21.53& 3950 &69.2±14.7 &73±0.5 &94.4±9.7 &94.4±9.7 &2.73±0.25 &1.36±0.24 &3.02±0.19 &3.02±0.19 &3.93 &5.21±0.29&4.76\\
6 &17:46:47.01, -28:32:08.88 &0.25& 1150 &- &- &- &- &- &-&-&-&0.79&-&-\\
7 &17:46:46.93, -28:32:08.57 &0.2& 1030 &95.4±34.6 &70±5.5 &- &95.4±34.6 &1.34±0.19 &1.34±0.42&-&1.34±0.19&0.05&2.09±0.38&2.31\\
8 &17:46:47.07, -28:32:07.26 &152.95& 8160 &- &- &165.4±10.5 &165.4±10.5 &- &-&2.04±0.06&2.04±0.06&15.29&2.70±0.12&2.97\\
\enddata
\end{deluxetable*}

\startlongtable
\begin{deluxetable*}{@{\extracolsep{-8pt}}ccccccccccccccc}
\tabletypesize{\tiny}
\tablenum{E2}
\tablewidth{\textwidth}
\tablecaption{Properties of the cores in Sgr~C.}
\label{tab:Sgr C cloud table}
\tablehead{
\colhead{Core ID} &  \colhead{RA and Dec} & \colhead{Flux} & \colhead{$R$} & \colhead{$T_{\rm H_2CO}$} & \colhead{$T_{\rm H_2CO}$} & \colhead{$T_{\rm CH_3CN}$} & \colhead{$T$} & \colhead{$\sigma_{\rm H_2CO}$} & \colhead{$\sigma_{\rm H_2CO}$} & \colhead{$\sigma_{\rm CH_3CN}$} & \colhead{$\sigma$} & \colhead{$M_{\rm core}$} & \colhead{$\mathcal{M}$} &  \colhead{$\alpha_{\rm vir,p}$} \\
\colhead{} &  \colhead{} &  \colhead{density} & \colhead{} & \colhead{LTE} & \colhead{non-LTE} & \colhead{LTE} & \colhead{} & \colhead{LTE} &  \colhead{non-LTE} & \colhead{LTE} & \colhead{} & \colhead{} & \colhead{} \\
\colhead{} & \colhead{(J2000)}& \colhead{(mJy)} & \colhead{(AU)} & \colhead{(K)} & \colhead{(K)} & \colhead{(K)} & \colhead{(K)} & \colhead{(\kms{})} & \colhead{(\kms{})}& \colhead{(\kms{})}& \colhead{(\kms{})}& \colhead{(\msol{})}& \colhead{} & \colhead{}
}
\colnumbers
\startdata
1 &17:44:41.38, -29:28:29.90 &19.33& 4620 &125.1±20.3 &140±7 &- &125.1±20.3 &2.34±0.1 &2.18±0.32&-&2.34±0.1&4.79&3.43±0.18&0.54\\
2 &17:44:41.32, -29:28:30.03 &0.34& 1360 &88±28 &101±63 &- &<89.1 &1.56±0.2 &1.27±0.37&-&1.56±0.2&<0.09&2.61±0.41&0.26\\
3 &17:44:40.81, -29:28:27.70 &0.37& 1290 &- &- &- &- &- &-&-&-&0.84&-&-\\
4 &17:44:41.21, -29:28:26.84 &1.1& 2330 &113.5±42.8 &125±16 &- &113.5±42.8 &1.56±0.18 &1.43±0.53&-&1.56±0.18&0.31&2.30±0.33&0.13\\
5 &17:44:41.20, -29:28:23.94 &0.61& 1680 &43.8±7.4 &65±39 &- &43.5±7.3 &1.24±0.14 &1.02±0.28&-&1.24±0.14&0.50&3.02±0.41&0.31\\
6 &17:44:40.76, -29:28:19.77 &0.36& 1370 &- &- &- &- &- &-&-&-&0.77&-&-\\
7 &17:44:40.79, -29:28:19.50 &1.15& 1310 &26.8±7.1 &26±26 &- &<89.9 &1.36±0.21 &1.35±0.31&-&1.36±0.21&<0.32&4.04±0.78&0.52\\
8 &17:44:40.78, -29:28:19.09 &1.14& 1310 &73.7±31.2 &130±12 &- &<99.5 &1.7±0.33 &1.53±0.63&-&1.7±0.33&<0.28&3.15±0.72&0.35\\
\enddata
\end{deluxetable*}

\startlongtable
\begin{deluxetable*}{@{\extracolsep{-8pt}}ccccccccccccccc}
\tabletypesize{\tiny}
\tablenum{E3}
\tablewidth{\textwidth}
\tablecaption{Properties of the cores in the 20~\kms{} cloud.}
\label{tab:20km table}
\tablehead{
\colhead{Core ID} &  \colhead{RA and Dec} & \colhead{Flux} & \colhead{$R$} & \colhead{$T_{\rm H_2CO}$} & \colhead{$T_{\rm H_2CO}$} & \colhead{$T_{\rm CH_3CN}$} & \colhead{$T$} & \colhead{$\sigma_{\rm H_2CO}$} & \colhead{$\sigma_{\rm H_2CO}$} & \colhead{$\sigma_{\rm CH_3CN}$} & \colhead{$\sigma$} & \colhead{$M_{\rm core}$} & \colhead{$\mathcal{M}$} &  \colhead{$\alpha_{\rm vir,p}$} \\
\colhead{} &  \colhead{} &  \colhead{density} & \colhead{} & \colhead{LTE} & \colhead{non-LTE} & \colhead{LTE} & \colhead{} & \colhead{LTE} &  \colhead{non-LTE} & \colhead{LTE} & \colhead{} & \colhead{} & \colhead{} \\
\colhead{} & \colhead{(J2000)}& \colhead{(mJy)} & \colhead{(AU)} & \colhead{(K)} & \colhead{(K)} & \colhead{(K)} & \colhead{(K)} & \colhead{(\kms{})} & \colhead{(\kms{})}& \colhead{(\kms{})}& \colhead{(\kms{})}& \colhead{(\msol{})}& \colhead{} & \colhead{}
}
\colnumbers
\startdata
1 &17:45:36.41, -29:06:30.01 &0.4& 1290 &- &- &- &- &- &-&-&-&0.71&-&-\\
2 &17:45:36.53, -29:06:29.63 &0.27& 1040 &- &- &- &- &- &-&-&-&0.36&-&-\\
3 &17:45:36.53, -29:06:29.38 &0.33& 1240 &- &- &- &- &- &-&-&-&0.51&-&-\\
4 &17:45:36.23, -29:06:29.28 &0.34& 1330 &- &- &- &- &- &-&-&-&0.55&-&-\\
5 &17:45:36.27, -29:06:28.57 &0.25& 1360 &- &- &- &- &- &-&-&-&0.52&-&-\\
6 &17:45:36.64, -29:06:26.83 &0.86& 1720 &101.5±49.7 &125±61 &- &101.5±49.7 &2.39±0.37 &2.99±0.96&-&2.39±0.37&0.21&3.90±0.70&6.12\\
7 &17:45:36.34, -29:06:26.57 &0.26& 1200 &- &- &- &- &- &-&-&-&0.42&-&-\\
8 &17:45:36.70, -29:06:24.69 &0.5& 1790 &61.2±20.4 &69.1±3.1 &- &61.2±20.4 &3.43±0.51 &3±1&-&3.43±0.51&0.21&7.34±1.18&11.18\\
\enddata
\end{deluxetable*}

\bibliography{H2CO}{}
\bibliographystyle{aasjournal}

\end{CJK}
\end{document}